\documentclass[sigconf]{acmart}

\AtBeginDocument{%
  }

\copyrightyear{2026}
\acmYear{2026}
\setcopyright{cc}
\setcctype{by}
\acmDOI{10.1145/3772318.3791061}

\acmConference[CHI '26]{Proceedings of the 2026 CHI Conference on Human Factors in Computing Systems}{April 13--17, 2026}{Barcelona, Spain}
\acmBooktitle{Proceedings of the 2026 CHI Conference on Human Factors in Computing Systems (CHI '26), April 13--17, 2026, Barcelona, Spain}

\acmISBN{979-8-4007-2278-3/2026/04}

\acmSubmissionID{1446}

\usepackage{soul}
\usepackage{bbding}

\newcommand{\parabox}[1]{%
  \begin{center}
    \setlength{\fboxsep}{6pt}%
    \fcolorbox{black}{gray!8}{%
      \parbox{0.9\linewidth}{#1}%
    }%
  \end{center}
  \par
}

\begin{document}

\title[Challenges and Opportunities in Supporting Version Control in Modern CAD Design]{Untangling the Timeline: Challenges and Opportunities in Supporting Version Control in Modern Computer-Aided Design}

\author{Yuanzhe Deng}
\affiliation{%
  \department{Mechanical and Industrial Engineering}
  \institution{University of Toronto}
  \city{Toronto, Ontario}
  \country{Canada}
}
\email{yuanzhe.deng@mail.utoronto.ca}

\author{Shutong Zhang}
\affiliation{%
  \department{Electrical and Computer Engineering}
  \institution{University of Toronto}
  \city{Toronto, Ontario}
  \country{Canada}
}
\affiliation{%
  \department{Computer Science}
  \institution{Stanford University}
  \city{Palo Alto, California}
  \country{USA}
}
\email{tonyzst@stanford.edu}

\author{Kathy Cheng}
\affiliation{%
  \department{Mechanical and Industrial Engineering}
  \institution{University of Toronto}
  \city{Toronto, Ontario}
  \country{Canada}
}
\email{kathy.cheng@mail.utoronto.ca}

\author{Alison Olechowski}
\affiliation{%
  \department{Mechanical and Industrial Engineering}
  \institution{University of Toronto}
  \city{Toronto, Ontario}
  \country{Canada}
}
\email{a.olechowski@utoronto.ca}

\author{Shurui Zhou}
\affiliation{%
  \department{ECE\&CS}
  \institution{University of Toronto}
  \city{Toronto, Ontario}
  \country{Canada}
}
\email{shurui.zhou@utoronto.ca}


\begin{abstract}
Version control is critical in mechanical computer-aided design (CAD) to enable traceability, manage product variation, and support collaboration. Yet, its implementation in modern CAD software as an essential information infrastructure for product development remains plagued by issues due to the complexity and interdependence of design data. This paper presents a systematic review of user-reported challenges with version control in modern CAD tools. Analyzing 170 online forum threads, we identify recurring socio-technical issues that span the management, continuity, scope, and distribution of versions. Our findings inform a broader reflection on how version control should be designed and improved for CAD and motivate opportunities for tools and mechanisms that better support articulation work, facilitate cross-boundary collaboration, and operate with infrastructural reflexivity. This study offers actionable insights for CAD software providers and highlights opportunities for researchers to rethink version control. 
\end{abstract}




\maketitle

\section{Introduction}
\label{sec:introduction}

With the growing complexity of engineering design, modern hardware product development increasingly depends on computer-aided design (CAD) software. 
With mechanical CAD, designers model the 3D geometric details of products with parametric commands and constraints, assemble parts for analysis, iterate designs, and deliver finalized specifications to downstream manufacturers~\cite{stark_major_2022}. 
While CAD originated as a digital substitute for pencil-and-paper drafting, it has since evolved into a virtual design studio where geographically distributed engineers discuss, argue, negotiate, and form consensus around design decisions~\cite{robertson_cad_1993, mitchell_cad_1995}. 
As CAD projects scale, they exhibit the hallmarks of a complex system as Boy defines~\cite{boy_human-centered_2017}: intricate interdependencies among components~\cite{cheng_its_2025}, increasingly distributed and collaborative workflows~\cite{cheng_age_2023}, sophisticated data management tooling~\cite{tang_research_2015}, and the nonlinear design processes resulting from parallel development, which must be reconciled~\cite{cheng_user_2023}. 
CAD has thus become not merely a technical drafting tool built for individual engineers but a complex socio-technical system at the centre of modern collaborative engineering and making. 

Throughout the product development process, design iterations are fundamental to exploring the design space and resolving failures, and the frequency of this iteration is strongly tied to design performance~\cite{vrolijk_connecting_2023, wynn_perspectives_2017}. 
Yet, these iterations generate rapidly expanding version histories that demand effective data management, capable of handling the complex interdependent CAD artifacts~\cite{kasik_ten_2005, frazelle_new_2021}. At the same time, collaboration across organizational and system boundaries further amplifies the system complexity~\cite{carayon_human_2006}, especially as each CAD system operates with different proprietary file formats~\cite{kasik_ten_2005, ji_machine_1997}. 
Consequently, version control has become an essential information infrastructure in CAD that structures relations among design versions, coordinates distributed work, and provides the context for effective communication~\cite{star_steps_1996, hanseth_developing_1996, star_steps_1994, cheng_age_2023}. 
Like any infrastructure, it remains largely invisible to the users until breakdown~\cite{star_steps_1996}, where such breakdowns may lead to consequences such as downstream manufacturing errors due to poor traceability~\cite{jarratt_engineering_2011}, loss of functionality due to broken component dependencies~\cite{ahmed_version_1991}, and chaotic coordination triggered by mismanaged branches of parallel development~\cite{cheng_user_2023}. 
These breakdowns hence motivate a systematic examination of the socio-technical challenges of modern CAD collaboration from the perspective of end users, with a particular focus on version control as an information infrastructure. 

While project failures and collaboration inefficiencies rarely stem from purely technical faults, they often arise from misalignment between technology and the social and organizational complexity of the environment in which it is deployed~\cite{baxter_socio-technical_2011}. 
For example, version control in CAD is traditionally implemented to be centralized and file-based, which requires users to download files, make changes locally, and then re-upload new versions to a central server. While this structure effectively minimizes editing conflicts, it restricts parallel work and slows distributed collaboration~\cite{cheng_age_2023}. 
However, as new CAD platforms such as \textsc{Onshape} enable real-time co-editing to mitigate these bottlenecks~\cite{baran_under_2015}, studies suggest that synchronous collaboration may increase communication overhead and coordination burdens~\cite{phadnis_are_2021, asuzu_personas_2024}. The optimal implementation is non-obvious, and ongoing studies are necessary.
Despite growing research on version control in CAD (e.g., traceability~\cite{cheng_age_2023, voigt_not_2018}, branching and merging~\cite{sun_method_2011, cheng_user_2023}, synchronous collaboration~\cite{chang_study_2007, asuzu_personas_2024, phadnis_are_2021}), there lacks a systematic understanding of the user challenges in version control, especially across the diverse landscape of modern CAD platforms~\cite{kannan_multi-criteria_2008}.
Without such an understanding, it is difficult to derive design principles for next-generation CAD infrastructures. 

Despite known technical limitations, users routinely restructure their workflows around existing technology, and infrastructures cannot be understood or improved without attending to these creative and improvisational practices of ordinary users~\cite{pipek_infrastructuring_2009}. 
For instance, consider the hobbyist open-source hardware (OSH) community, where contributors frequently share 3D printing-ready CAD models in diverse proprietary formats on platforms such as \textsc{Thingiverse}, which provide little to no version control support~\cite{cheng_lot_2024, ludwig_towards_2014}. 
To compensate for this infrastructural limitation in tracking design versions, OSH hobbyists repurpose the model description field to document change history, despite the space being intended for functional summaries and 3D printing instructions~\cite{cheng_lot_2024}. 
Bottom-up workarounds like this highlight the gap between formal infrastructure and actual practice; this study surfaces and synthesizes the real-world challenges users encounter as well as their creative workarounds when performing version control in CAD. 

In this paper, we address several important socio-technical questions to establish a foundation from which to innovate: Where does current CAD infrastructure fall short in supporting collaborative design; how do users navigate fragmented versioning workflows; and how might future CAD systems better support traceability, design variation, and distributed coordination? 
To answer these questions, we conducted a qualitative analysis of 170 version-control-related user posts and their associated discussions, collected from seven CAD-related online forums. 
Through inductive content analysis, we identified four high-level categories of version control challenges, concerning the management, continuity, scope, and distribution of versions. 
Building on these findings, we derive three high-level design opportunities for advancing version control support in CAD, including better supporting articulation work throughout the design process, facilitating cross-boundary collaboration, and developing version control that foregrounds reflexivity as a key design principle. 
Together, these insights re-emphasize CAD not merely as a technical tool but as a socio-technical system whose technical development must be shaped by the practical realities of its use. 

In summary, we contribute the following insights to the growing CAD community within the field of human-computer interaction (HCI): 
\begin{itemize}
    \item An empirical investigation of the user challenges in performing version control with CAD, based on public discussions among users across seven online forums. 
    \item A comparison of functionalities across three commercially available CAD software platforms in support of the identified version control challenges.
    \item A set of design opportunities for version control in CAD as an information infrastructure. 
\end{itemize}

\section{Background and Related Work}

Originating in software engineering~\cite{chou_unifying_1986, ruparelia_history_2010}, the concept of version control has since expanded to domains such as collaborative creative practices~\cite{sterman_towards_2022}, scientific workflow management~\cite{abediniala_facilitating_2022}, and modern engineering design in CAD. 
In this section, we review the types of version control system (VCS) used in CAD (Section~\ref{sec:vc_type}) and their applications (Section~\ref{sec:vc_use}), followed by the research gap in Section~\ref{sec:bkg_gap}. 

\subsection{Types of Version Control}\label{sec:vc_type}

The complexities of CAD as a design platform result in particularly vexing challenges~\cite{ahmed_version_1991, chou_unifying_1986, van_den_hamer_managing_1996}, and have resulted in varied implementations of version control. 

\subsubsection{Centralized version control} 
\label{sec:bkg_cvcs}

Traditional version control in CAD is largely supported through a \emph{local centralized version control system} (CVCS), often via product data management (PDM) or product lifecycle management (PLM) systems~\cite{stark_pdm_2016, liu_design_1990, chou_unifying_1986}, which originated from software development tools like \textsc{CVS}\footnote{https://cvs.nongnu.org} and \textsc{Subversion}\footnote{https://subversion.apache.org}. In this model, a central server maintains the complete version history, and users acquire access to the versions as a client through network connection~\cite{ruparelia_history_2010}. 
To modify a CAD file, users must first retrieve the file from the server (i.e., ``checking out'' the model) to their local machine with the CAD software installed, serving to lock the file and preventing concurrent edits. When complete, the user ``checks back in'' the updated version to the server. 

As cloud infrastructure matures, PDM systems have also transitioned to \emph{cloud-based CVCS}~\cite{heventhal_fusion_2018}, enabling distributed access to CAD model data and eliminating the overhead cost of infrastructure setup~\cite{junk_use_2018}. This shift facilitates remote collaboration across geographic locations~\cite{barrie_applications_2016}. 
However, the fundamental version control workflow remains centralized, where copies of CAD files must be checked out, modified locally, and uploaded in full after every version update. 

Although the check-in/check-out workflow in CVCS avoids editing conflicts, it introduces delays and limits concurrent editing. Users often only become aware of design changes after a file is checked in, sometimes leading to integration issues when interdependent components are developed in parallel~\cite{cheng_age_2023, bicici_collaborative_2012}. 
Despite these limitations, CVCS remains the dominant approach in modern CAD, especially when exchanging models across different CAD platforms or in open-source environments ~\cite{cheng_lot_2024}. 
While prior research has documented some inefficiencies in CVCS workflows, a systematic understanding of how these technical constraints impact user collaboration is lacking. In this paper, we review the challenges from the users' perspectives and examine how they adapt to these limitations in practice. 

\subsubsection{Distributed version control} 
\label{sec:bkg_dvcs}

In contrast, distributed version control systems (DVCS) like \textsc{Git}\footnote{https://git-scm.com} allow users to work independently by maintaining isolated local repositories, supporting parallel development and more flexible workflows. DVCS tracks individual changes at a finer granularity (e.g., line-level differences), resulting in smaller and more cohesive commits with improved traceability~\cite{brindescu_how_2014}. 
Modern CAD software achieves \emph{cloud-based DVCS} by migrating both the CAD software and data to the cloud~\cite{frazelle_new_2021}. This architecture enables users to access modelling tools directly through a web browser, where every design action is continuously synchronized with the cloud. 
Rather than uploading whole files, changes are incrementally recorded and integrated in real-time, eliminating manual check-in/check-out~\cite{onshape_eliminating_2024, la_fleche_gitflow_2019}.
These systems support simultaneous multi-user collaboration on the same model, enabling a new class of cloud CAD or multi-user CAD workflows~\cite{baran_under_2015}. 

Despite these advances, the implications of DVCS in CAD remain under-explored~\cite{french_collaborative_2016, deng_does_2022}. While DVCS resolves key limitations of CVCS (i.e., low concurrency in collaboration), it introduces new challenges.  
For instance, Cheng et al. observed that synchronous editing can result in design conflicts and duplicate work due to insufficient awareness of teammates' actions~\cite{cheng_analysis_2024}. 
Besides, most existing studies on cloud CAD focus on examining design quality through surface-level metrics such as user satisfaction, communication patterns, and telemetric action data~\cite{phadnis_are_2021, deng_multi-user_2022, eves_comparative_2018, phadnis_multimodal_2021, zhou_analysis_2021}, rather than analyzing how version control mechanisms influence collaborative workflows. 
In this paper, we systematically examine the user-facing challenges across both centralized and distributed VCS, and we discuss the implications for future CAD software providers. 

\subsection{Use of Version Control}\label{sec:vc_use}

In order to review the uses of version control in CAD, we apply Estublier and Casallas' three orthogonal classes of versioning, established in the context of software engineering: historical, logical, and cooperative, each with different motivations and circumstances~\cite{goos_three_1995}. 

\subsubsection{Historical versioning for traceability} 
\label{sec:bkg_trace}

Historical versioning records the linear progression of a project in the time dimension, enabling traceability and accountability~\cite{goos_three_1995, watkins_why_1994}. In CAD, VCS captures design milestones as immutable snapshots, preserving the evolution of a model throughout its lifecycle~\cite{krishnamurthy_version_1995}. These versions provide essential backup points when fixing modelling errors and facilitate design reuse when revisited. 
Ideally, these snapshots are semantically meaningful representations of the design history~\cite{chou_unifying_1986, hofmann_greater_2018}. 

Maintaining high-quality traceability is resource-intensive in both software~\cite{maro_traceability_2016, lago_scoped_2009} and CAD~\cite{cheng_age_2023}, particularly in large-scale projects with complex interdependencies. In CAD, traceability is further complicated by the binary and proprietary nature of the model files, which hinders efficient comparison between versions~\cite{cheng_age_2023}. 
OSH communities, where contributors frequently use different tools and platforms, face additional challenges due to increased communication overhead and inconsistent version tracking practices~\cite{cheng_lot_2024}. 
Although the importance of traceability is well-established in case studies~\cite{cheng_lot_2024} and technology review articles~\cite{kasik_ten_2005, piegl_ten_2005, frazelle_new_2021}, the specific burdens and breakdowns encountered by general CAD users remain under-analyzed. This motivates a closer investigation into how different types of VCSs (as reviewed in Section~\ref{sec:vc_type}) support or hinder traceability in real-world CAD workflows. 

\subsubsection{Logical versioning for variation} 
\label{sec:bkg_branch}

Logical versioning supports the coexistence of multiple variants of a design, each representing a parallel evolution path of the project~\cite{goos_three_1995}. In software engineering, this involves \emph{branching} (creating copies of) the source file for isolated development and \emph{merging} (integrating) changes back into a target branch~\cite{baudis_current_2014, premraj_branch_2011}. In CAD, branching is used mainly to manage product lines (e.g., product releases, design variants), isolate development risk, and coordinate work among collaborators~\cite{cheng_user_2023}. 

In CAD, managing variants becomes increasingly complex as the number of branches grows~\cite{cheng_user_2023}, as is the case in  software~\cite{linsbauer_classification_2017, phillips_branching_2011}. Branching in CAD introduces unique challenges due to the tightly coupled geometric and parametric dependencies among models. 
Unlike in text-based source code, where developers can selectively merge specific changes (i.e., cherry-picking\footnote{https://git-scm.com/docs/git-cherry-pick}), branching in CAD typically results in entire model variants being duplicated, without fine-grained control over which features or modifications are preserved, discarded, or merged~\cite{cheng_user_2023}. 
Although recent advancements in cloud CAD have made branching and merging more accessible, this functionality also introduces new coordination demands. In practice, an ``integrator'' role often emerges, responsible for resolving merging conflicts and maintaining up-to-date model versions across branches~\cite{asuzu_personas_2024}. 
Extending beyond Cheng et al.'s review~\cite{cheng_user_2023} on the use cases and deficiencies of branching in CAD, this paper systematically reviews how both branching and merging are supported and hindered in practice. 

\subsubsection{Cooperative versioning for synchronization} 
\label{sec:bkg_sync}

Cooperative versioning supports real-time collaboration by synchronizing changes across users~\cite{mikkonen_elements_2012}. 
Cloud-based CAD systems automatically create implicit versions when changes are detected, enabling synchronous updates across devices and locations. These automatically created implicit versions complement the explicit versions used for historical and logical purposes, offering a layered versioning model that balances flexibility and control. 
As discussed in Section~\ref{sec:bkg_dvcs}, cloud-based DVCS eliminates the check-in/check-out bottleneck of traditional CVCS and enables real-time multi-user editing. However, this increased synchronicity introduces new challenges, and a comprehensive review is lacking in understanding the collaborative trade-offs introduced by these advances in VCS. 

\subsection{Fundamental Differences Between CAD and Software Version Control}
\label{sec:bkg_diff_field}

Although version control originates in software engineering, the challenges of managing, interpreting, and coordinating versions in CAD cannot be captured by existing software-centric VCS models. From an HCI perspective, CAD introduces fundamentally different forms of representational friction and socio-technical limitations that reshape how people collaborate and make sense of evolving artifacts: 
\begin{itemize}
    \item \emph{Geometric complexity.} CAD models are built through sequences of parametric operations and constraint networks that implicitly encode design intent: a single dimension change may cascade across multiple parts and assemblies~\cite{cheng_its_2025, camba_parametric_2016, piegl_ten_2005}. This makes various sophisticated operations that are fundamental in software VCS (e.g., diff, merge) not merely unsupported in traditional CAD systems but conceptually misaligned with how CAD artifacts behave in their current forms~\cite{frazelle_new_2021, briere-cote_3d_2013, masmoudi_dependency_2015}. 
    \item \emph{Versioning workflows.} While traditional engineering change management practices document detailed justifications of each revision~\cite{wright_review_1997, jarratt_engineering_2011}, modern cloud-based CAD systems streamline the process and generate implicit versions continuously to enable synchronous collaboration, creating dense but low-semantic version histories~\cite{baran_under_2015}. Unlike text-based code versions that can be queried, compared, and navigated, CAD versions require users to reconstruct the design intent through geometric inspection and temporal inference. Without this labour, the version history quickly loses its meaning and becomes impossible to navigate~\cite{cheng_age_2023}. 
    \item \emph{Collaboration span.} CAD assemblies rely on references to components that span multiple files and off-the-shelf parts imported from other platforms~\cite{cheng_its_2025, chu_multi-skeleton_2016}. However, when CAD models are transferred across projects or exported to lightweight neutral formats, parametric information and design history are often stripped entirely to reduce the file size, leaving only ``dumb'' geometries that are not editable~\cite{kasik_ten_2005}. These breakdowns do not manifest as textual conflicts in source code, and therefore the ways users negotiate meaning, ownership, and repair in CAD extend beyond what software VCS research typically examines. 
\end{itemize}
Taken together, these representational complexities and collaborative demands show that CAD VCS cannot be understood as a simple instance of software VCS in another domain.
Instead, it constitutes a distinct class of HCI challenges involving cross-representation sensemaking, coordination around interdependent artifacts, and the management of complex geometry-based design histories. These differences motivate the need for a user-centred investigation of how people experience and work around CAD-specific version control breakdowns, and how future systems might better support CAD and other complex design workflows.

\subsection{Research Gap}
\label{sec:bkg_gap}

Despite the central role of version control in collaborative design, existing research on CAD versioning remains fragmented. Prior work has largely focused on technical implementations of VCS or has studied collaboration in CAD without in-depth analysis of the supports and limitations of the VCS. 
There remains a lack of systematic cross-platform understanding of the socio-technical challenges users face when applying version control in real-world collaborative practices in CAD.
As the product design community acknowledges the importance of version control during product development, and an increasing number of businesses have integrated the use of version control in their daily design activities, a review of user feedback is required.
This study addresses that gap by synthesizing user-reported experiences and empirical findings to examine where current CAD version control workflows succeed, where they fall short, and what users need to support more effective and collaborative design. 

\section{Methodology}

In this section, we present an overview of our study scope and descriptions of our data collection and analysis methods. 

\subsection{Study Scope} 
\label{sec:study_scope}

In order to effectively understand the varying use cases and challenges in version control for mechanical CAD, we primarily focused on studying the users' online discussions from three model creation platforms (Section ~\ref{model-creation}), two model sharing platforms (Section ~\ref{model-sharing}), and two general online forums that are not officially affiliated with any commercial CAD software (Section ~\ref{online-forum}), as summarized in Table~\ref{tab:platforms}. 
This study expands the scope of a previous HCI study~\cite{cheng_user_2023}, which gained valuable insights on branching in CAD from a user perspective. 

We chose to analyze online forum discussions because they provide candid and naturally occurring user-reported challenges of real-world CAD tool use, surfacing organically in response to breakdowns, unmet needs, and design challenges. 
By examining the full discussion threads rather than isolated posts, we were able to reconstruct user intent, situational constraints, and problem-solving strategies with greater context and fidelity. 
This approach reveals not only users' perceptions of technical limitations, but also how users adapt, coordinate, and develop informal workarounds across a broad range of design contexts, offering ecologically valid insights into infrastructural friction, collaborative dynamics, and tool appropriation across distributed design workflows. 

\begin{table*}[ht]
    \centering
    \caption{Platforms selected for the study of version control in CAD.}
    \renewcommand{\arraystretch}{1.3}
    \begin{tabular}{l l l l l l}
        \toprule  
        \textbf{Type} & \textbf{Platform} & \textbf{Launch} & \textbf{Intro. of VCS} & \textbf{VCS Type} & \textbf{Data Source} \\
        \midrule  
         Model creation  & \textsc{SolidWorks} & 1995 & 2003 & Local CVCS & \textsc{SolidWorks} Forums\textsuperscript{a} \\
        & \textsc{Autodesk Fusion} & 2013 & 2013 & Cloud CVCS & \textsc{Autodesk} Forums\textsuperscript{b} \\
        & \textsc{Onshape} & 2015 & 2015 & Cloud DVCS & \textsc{Onshape} Forums\textsuperscript{c} \\
        \midrule
         Model sharing & \textsc{Thingiverse} & 2008 & N/A & None & \textsc{Thingiverse} Groups\textsuperscript{d} \& \textsc{Reddit}\textsuperscript{e} \\
        & \textsc{Thangs} & 2020 & 2020 & CVCS & \textsc{Reddit}\textsuperscript{e} \\
        \midrule
         General forum & \textsc{Eng-Tips} & 1997 & N/A & N/A & \textsc{Eng-Tips}\textsuperscript{f} \\
        & \textsc{CAD Forum} & 2021 & N/A & N/A & \textsc{CAD Forum}\textsuperscript{g} \\
        \bottomrule 
    \end{tabular}
    {\footnotesize
    \begin{tabular}{@{} p{0.8\linewidth} @{} @{} @{} @{}}
        \textsuperscript{a} https://forum.solidworks.com; 
        \textsuperscript{b} https://forums.autodesk.com; 
        \textsuperscript{c} https://forum.onshape.com; 
        \textsuperscript{d} https://www.thingiverse.com/groups/thingiverse; 
        \textsuperscript{e} https://www.reddit.com; 
        \textsuperscript{f} https://www.eng-tips.com/; 
        \textsuperscript{g} https://cadforum.net/ and later migrated to https://forum.cadmunity.com/ in May 2025.
    \end{tabular}}
    \label{tab:platforms}
\end{table*}

\subsubsection{Model Creation Platforms}\label{model-creation}

For product development, the official user forums of three commercially available CAD platforms were included in our study (\textsc{Autodesk Fusion}, \textsc{Onshape}, and \textsc{SolidWorks}). These are also the only three CAD software platforms that enable built-in version control for 3D CAD models~\cite{heventhal_fusion_2018, onshape_eliminating_2024, noauthor_solidworks_2019}, each representing one type of VCS introduced in Section~\ref{sec:vc_type}, as shown in Table~\ref{tab:platforms}. In other CAD software, such as \textsc{OpenSCAD}, \textsc{Creo}, and \textsc{Autodesk Inventor}, enabling version control requires the purchase of external PDM/PLM solutions. At the same time, the selected CAD platforms not only target professional engineering designers for commercial use, but also provide different kinds of account types for design hobbyists with reduced support of advanced features. 

All three selected CAD modelling platforms host an online forum for users to post and discuss product-related topics (e.g. feature-related questions and improvement requests) as a communication platform between the users and the software providers. These forums are listed in the last column of Table~\ref{tab:platforms}. Access to post and participate in these forums is limited to users with a registered account with the CAD software package, which ensures the relevance and quality of the discussion. 

\subsubsection{Model Sharing Platforms}\label{model-sharing}

CAD modelling is increasingly important for hobbyists, too, with the growing popularization of 3D printing~\cite{JUNK2016430}. Both \textsc{Thingiverse}\footnote{https://www.thingiverse.com} and \textsc{Thangs}\footnote{https://thangs.com} attempt to lower the barrier of entry to 3D printing by providing a platform for hobbyists to quickly search for and use CAD models that suit their needs. 

Most notably, \textsc{Thingiverse} has built the largest maker community for sharing 3D-printable CAD models, with over 3 million models exchanged between over 2.3 million registered users~\cite{leighton_makerbot_2018}, making it a major platform for studies in collaborative open product development~\cite{cheng_lot_2024, alcock_barriers_2016, buehler_sharing_2015}. In \textsc{Thingiverse}, user discussions about the platform take place in different Groups, and we focused our study on the ``General'' forum of the ``Thingiverse'' group, the official Group managed by the platform.

We chose to include \textsc{Thangs} in this study, as it is the first to support the search for similar models across different CAD model-sharing platforms (e.g., \textsc{Thingiverse}, \textsc{Free3D}), a version-control-related feature that is widely absent from other model-sharing platforms. Based on a user-uploaded model, \textsc{Thangs} will attempt to locate different versions of the same design~\cite{physna_introducing_2020} and provide a version control workspace to visualize the geometric difference between CAD models~\cite{thangs_3d_thangs_2022}. \textsc{Thangs} does not have an officially affiliated forum for discussion, but there is an official user account that is active on \textsc{Reddit},\footnote{https://www.reddit.com/user/Thangs3D/} an online forum with communities organized by topic. To collect discussions around version control on \textsc{Thangs}, we searched four subreddit communities where the official \textsc{Thangs} account actively participates: ``r/3Dprinting'', ``r/resinprinting'', ``r/Thangs3D'', and ``r/blender''. 

\subsubsection{Online Forums}\label{online-forum}

Beyond gathering user discussions from product-specific forums, we examined two general CAD-related forums (\textsc{CAD Forum} and \textsc{Eng-Tips}) that are managed by third parties and are meant for users to freely share cross-platform knowledge and discussions. Here we expected posts on the generic usage of version control in CAD and comparisons of solutions between major CAD software providers. For the \textsc{Eng-Tips} forum, we specifically focused on two relevant sub-forums: ``Mechanical engineering general discussion'' under the ``Mechanical Engineers'' section and ``Engineering programs/apps (general) Forum'' under the ``Engineering Computer Programs'' section. 

\subsection{Data Collection}
\label{sec:data_collection}

\paragraph{Keyword search}
We developed the following set of keywords to use when searching for posts and discussions in the selected online forums: ``\emph{version control},'' ``\emph{change history,}'' ``\emph{versioning,}'' ``\emph{version history,}'' and ``\emph{variation management}.'' These keywords were selected because they either directly refer to the use and need for version control or relate to the major applications of version control in mechanical product design settings. 
Initial exploratory searches using broader terms such as the term ``version'' alone returned more than 20{,}000 threads, the vast majority of which were unrelated to CAD version control but rather discussions about software compatibility and licensing issues, for example. 
Using a more focused set of keywords allowed us to prioritize precision and ensure that the retrieved threads specifically addressed the version-control challenges examined in this study.
Nevertheless, to verify that this search strategy did not exclude major categories of discussion, we conducted iterative spot-checks using broader keyword queries (e.g., ``version'', ``history''). These checks did not reveal additional themes beyond those present in the final dataset, suggesting that the selected threads adequately captured the conceptual space of interest and that thematic saturation was reached.

\begin{figure*}
    \centering
    \includegraphics[width=0.9\linewidth]{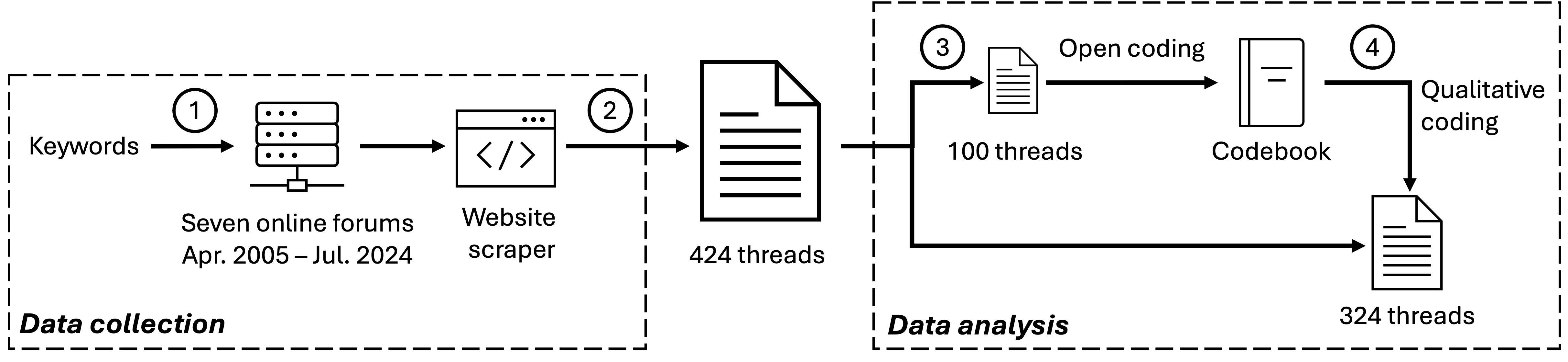}
    \caption{Overview of the research methodology. (1) A list of keywords was first defined in Section~\ref{sec:data_collection} to search for posts and discussions from seven forums. (2) A total of 424 threads from the forum search are mined with a self-developed website scraper. (3) A subset of 100 threads was first randomly selected for open coding with the method described in Section~\ref{sec:data_analysis}, and a coding scheme was developed. (4) The established codebook was then used to qualitatively code the remaining 324 threads.}
    \Description{The flowchart starts with the data collection phase, where a set of keywords were used to extract posts and discussions between April 2005 and July 2024 from seven online forums using a self-developed website scrapper. This phase resulted in 424 threads collected for analysis. Then, the data analysis phase contains two paths: the first path randomly selected 100 threads for open coding, and the developed codebook was used to qualitatively code the remaining 324 threads in the second path.}
    \label{fig:coding_methods}
\end{figure*}

\paragraph{Time window}
We defined a time window of relevance for the forum threads, which began April 2005 (when version control was introduced in software development by \textsc{Git}~\cite{brown_git_2018}) to July 2024 (when the data were collected). From the data sources presented in Table~\ref{tab:platforms}, the threads were collected via the workflow visualized in Figure~\ref{fig:coding_methods}.
To collect \textsc{Reddit} threads, we used the \textsc{Project Arctic Shift}.\footnote{https://github.com/ArthurHeitmann/arctic\_shift} For all other platforms, a self-developed web-scraper was built in Python using the \textsc{Beautiful Soup}\footnote{https://www.crummy.com/software/BeautifulSoup/} and \textsc{Selenium}\footnote{https://selenium-python.readthedocs.io} packages to collect all the textual information. 
The step resulted in a total of 424 forum threads.

\paragraph{Web scraping}
For each thread, we collected the title and content of the initial post, along with the user name of the author and the publishing date. We also collected all comments on those posts, along with the commentor's username. This approach allowed us to infer intent, situational constraints, and user goals with greater fidelity. The total number of threads collected from each platform is tabulated in Table~\ref{tab:posts_count}. 

\subsection{Data Analysis}
\label{sec:data_analysis}

\begin{figure*}
    \centering
    \includegraphics[width=\linewidth]{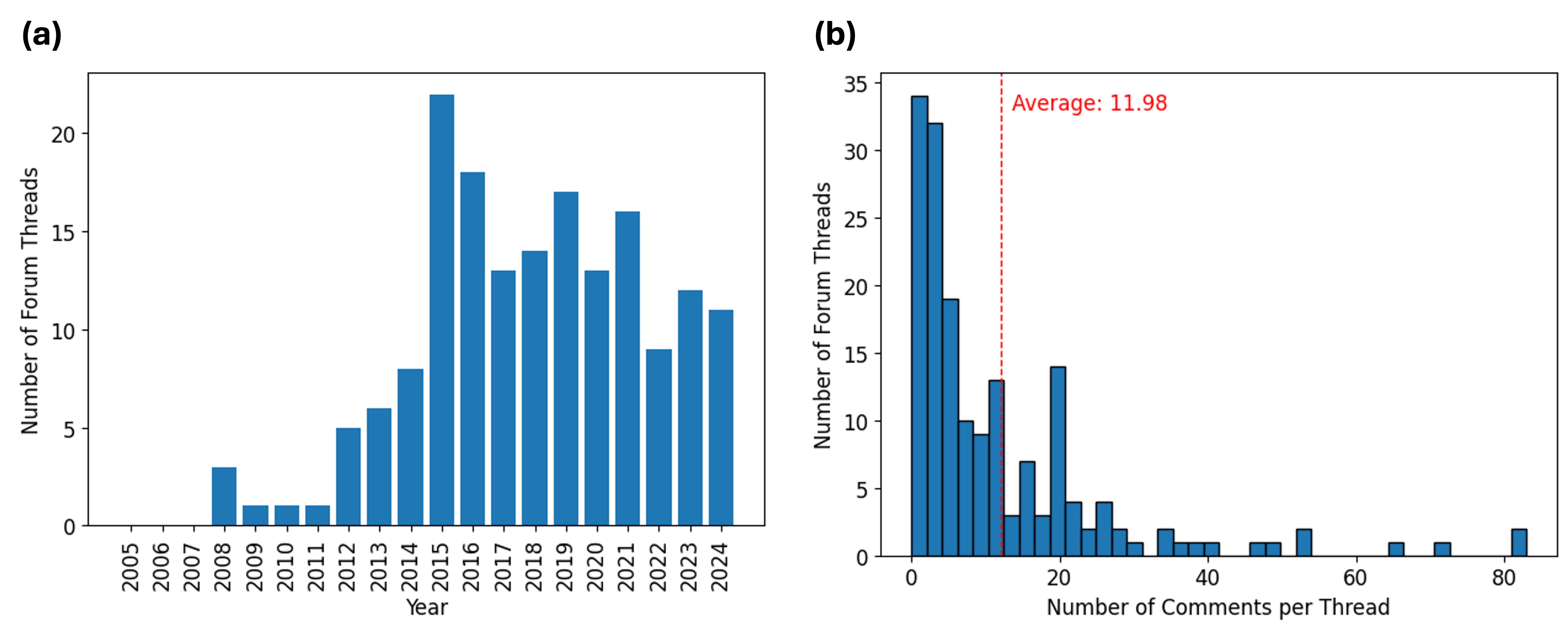}
    \caption{Characteristics of the dataset of 170 relevant forum threads included in our analysis. (a) Temporal distribution of forum threads from April 2005 to July 2024. (b) Distribution of comment counts per forum thread. On average, each thread contains 11.98 comments.}
    \label{fig:data_vis}
    \Description{In figure (a), a bar plot shows the number of relevant forum threads that were created in each year, where the first relevant thread appeared in 2008, and less than 10 relevant threads are collected per year between 2008 and 2014. After 2014, the number of relevant threads collected from each year ranges between 10 and 23. In figure (b), the distribution plot shows that the number of comments per thread ranges between 0 and over 80, where the average is 11.98. The majority of threads has less than 20 comments.}
\end{figure*}

Next, two authors of the paper manually coded the collected threads, as illustrated in Figure~\ref{fig:coding_methods}. One coder has expertise in version control for product design in CAD, and the other has expertise in version control for software development with \textsc{Git}. Following the Institute of Education Science (IES) guidelines~\cite{institute_of_education_science_reviewer_2017}, a subset of 100 threads ($> 20 \%$ of all threads collected) was first randomly selected for open coding~\cite{saldana_coding_2013}. The two coders began by screening the subset and excluding irrelevant threads (e.g., discussions on debugging of technical features). Based on the remaining subset, the coders developed a scheme to classify the relevant discussions into categories of user challenges, where the inter-assessor agreement (IAA) was calculated to be $83\%$ total agreement. With the obtained IAA exceeding the minimum threshold ($> 80\%$) proposed by the IES guidelines~\cite{institute_of_education_science_reviewer_2017}, the coders then proceeded to individually complete the analysis of the remaining dataset using the developed coding scheme, where each thread was first evaluated for relevancy and then assigned to one or more codes (i.e., user challenges), if appropriate. 
For the full dataset, the initial round of coding resulted in 81.4\% total agreement, 9.67\% partial agreement (two coders agreed on one or more codes assigned, but not all), and 8.96\% disagreement. The two coders discussed and resolved all disagreements from the initial round of coding. 
In total, 170 threads were identified as relevant to the use of version control in CAD, where the number of threads collected per forum is reported in Table~\ref{tab:posts_count}, and more characteristics of the dataset are visualized in Figure~\ref{fig:data_vis}.

\begin{table}[ht]
    \caption{Number of threads collected and analyzed from the selected data sources.}
    \label{tab:posts_count}
    \centering
    \begin{tabular}{l p{0.29\linewidth} p{0.28\linewidth}}
        \toprule 
        \textbf{Forums} & \textbf{Threads collected} & \textbf{Relevant threads identified} \\
        \midrule  
        \textsc{Autodesk} & 135 & 66 (49\%) \\
        \textsc{Onshape} & 136 & 55 (40\%) \\
        \textsc{SolidWorks} & 67 & 20 (30\%) \\
        \textsc{Reddit} & 49 & 17 (35\%) \\
        \textsc{CAD Forum} & 28 & 9 (32\%) \\
        \textsc{Eng-Tips} & 8 & 3 (38\%) \\
        \textsc{Thingiverse} Groups & 1 & 0 (0\%) \\
        \midrule
        \textbf{Total} & \textbf{424} & \textbf{170} \\
        \bottomrule
    \end{tabular}
\end{table}

\section{Results}
\label{sec:result}

\begin{figure*}
    \centering
    \includegraphics[width=0.85\linewidth]{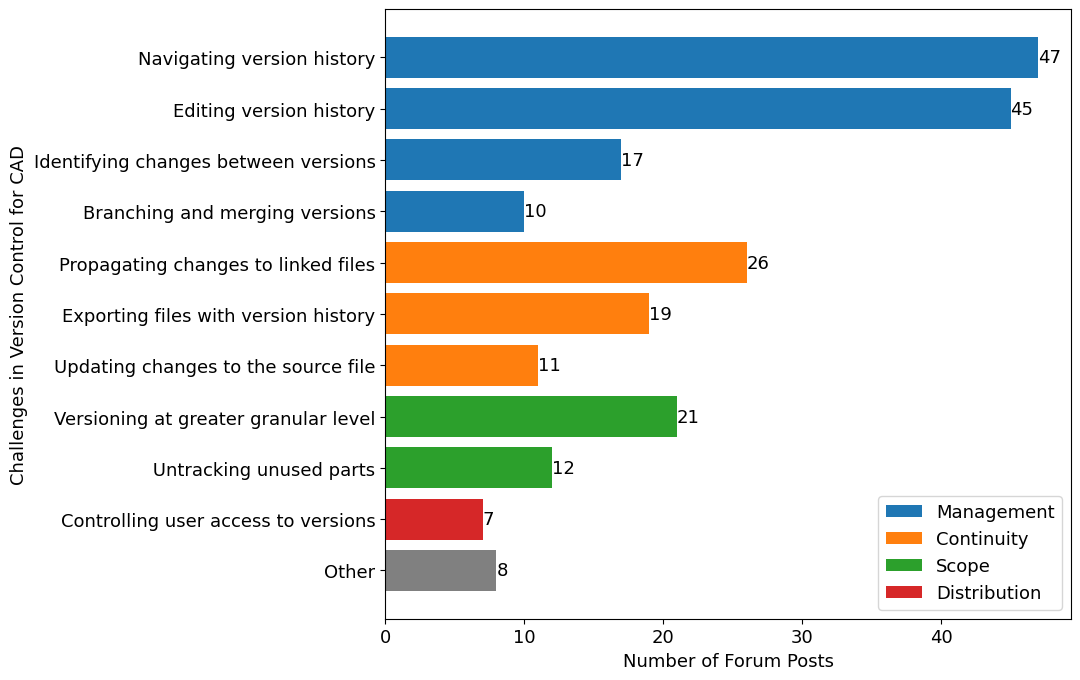}
    \caption{Frequency of challenges with using version control in CAD reported in the relevant forum threads ($n=$170), where one or more challenges were identified in each thread. Challenges are grouped into four categories: management, continuity, scope, and distribution of version control in CAD.}
    \Description{A bar graph that visualizes the frequency of user challenges reported in the dataset, and each user challenge is colour-coded based on the category that it falls under (i.e., one of the categories in Management, Continuity, Scope, and Distribution). The specific user challenge, the corresponding category, and the number of occurrences are all reported in this results section in detail.}
    \label{fig:coding_result}
\end{figure*}

Through open coding of all 170 relevant threads identified in Table~\ref{tab:posts_count}, we derived four categories of user challenges: management, continuity, scope, and distribution of versions in CAD, the results of which are summarized in Figure~\ref{fig:coding_result}. Challenges that were only reported in three threads or less were grouped as ``Other''. These categories were derived on the basis of patterns that emerged directly from the data, rather than predefined frameworks. 
In this section, we present the detailed descriptions and analysis of each coding category. Where excerpts from forum threads are quoted, we reference the source of the thread with the forum name and the posting year in brackets; for example, ``(OS21)'' refers to a thread from the \textsc{Onshape} forum that was posted in 2021.\footnote{Abbreviations for the seven forum sites are: \textsc{Autodesk} (AD), \textsc{Onshape} (OS), \textsc{SolidWorks} (SW), \textsc{Reddit} (RD), \textsc{Thingiverse} Groups (TV), \textsc{CAD Forum} (CF), and \textsc{Eng-Tips} (ET).} 

\subsection{Management of Version Control in CAD}

This section discusses the challenges users raised in managing the accelerating number of versions created, either automatically by the software or manually by the users, over the course of a product's development. 

\subsubsection{Navigating Version History}
\label{sec:navigate_ver}

In our study, the most frequently discussed challenge concerns the navigation of a model's version history, mentioned in 47 forum threads. 
CAD systems typically use an incremental numbering scheme (e.g., v1, v1.1) to trace design evolution, either assigned automatically or imposed by organizational protocols~\cite{stark_major_pdm_2022}. However, this naming scheme quickly becomes unmanageable, especially for CAD platforms with a CVCS. 
As one Autodesk user elaborates: ``\emph{In theory this works, but us humans get hammered into us to save regularly so we never lose work if there is a power failure. This means that we get way more versions than we actually need and annoying by the time it gets to V50 or more. By the time you get into these larger version numbers, the advantage of opening a specific one to go back in time becomes somewhat redundant as any one of several will do}'' (AD24). 
This reflects a common trend across forums: version numbers increase, but they convey little useful information.

\begin{figure*}
    \centering
    \includegraphics[width=1\linewidth]{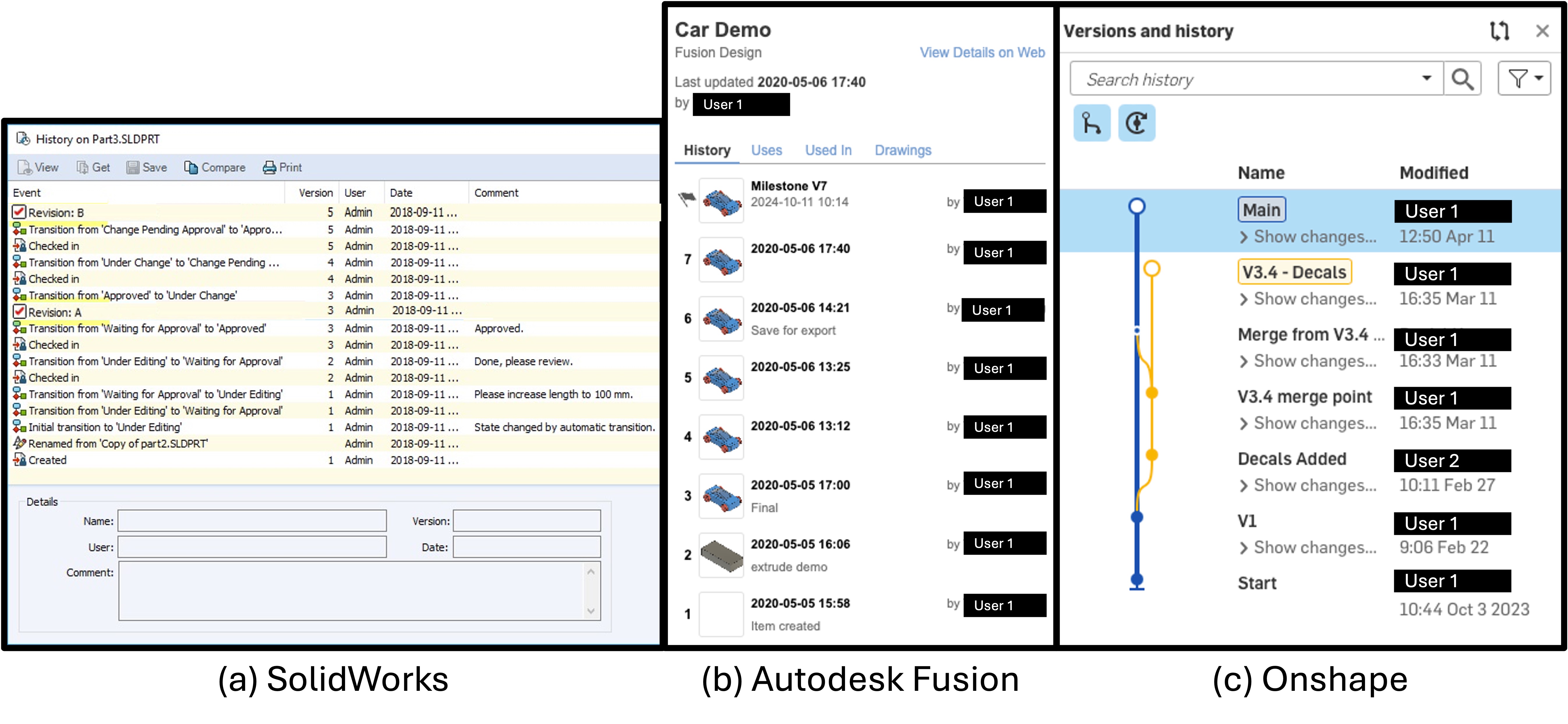}
    \caption{Display of version history of a file in the three studied CAD software. (a) \textsc{SolidWorks} automatically generates versioning event descriptions and incremented version numbers for every PDM action (e.g., check-in, check-out, revision), with the option for users to enter detailed descriptions (adapted from Ref.~\cite{lohmann_can_2018}). (b) \textsc{Autodesk Fusion} creates an automatically incremented version number for every file version saved and synced to the cloud. Unless the user otherwise provides a short description, only the timestamp is available to label the created versions. Users can also create milestones (based on the current version) with custom names and detailed descriptions, displayed with a flag sign. (c) \textsc{Onshape} cascades all changes to the file under every user-created version, and branches are shown with additional visualization (i.e., a yellow branching path from the main blue branch). All three platforms display the user-specified version description with limited space in the history view, where additional clicks are required to open a separate dialogue window for a specific version to access the detailed documentation. All screenshots of software interfaces are presented for comparison purposes only.}
    \Description{This figure presents the presentation of version history in the three commercial CAD modelling platforms that we included in this study as a comparison. These commercial CAD software presents version histories as a pop-up window or a side-panel. Key characteristics of the visualization are summarized in the figure caption.}
    \label{fig:version_view}
\end{figure*}

Beyond their sheer volume, numeric labels lack descriptive summaries, offering little insight into the content or purpose of each version. As a result, version timelines become difficult to navigate, particularly with limited access to detailed documentation, as illustrated in Figure~\ref{fig:version_view}. 
As one user stated, sorting through versions ``\emph{will become a nightmare and leads to this problem of having to navigate the document history to exactly where and if something was released}'' (OS19), highlighting the navigational challenges for both designers and downstream manufacturers. 
Across all three tools shown in Figure~\ref{fig:version_view}, the interfaces also compress descriptions into narrow columns, obscuring contextual information unless users manually expand them with additional clicks. Although this interface design enables a large volume of information to be displayed within a compact view, the resulting loss of visibility slows down both design review and version retrieval processes. 

Even when versions include user-written descriptions, searching for a specific design modification remains difficult. For instance, an \textsc{Onshape} user described that ``\emph{instead of reading through 1000's of lines and sorting through dozens of hastily made versions, I just want to find the last time I was working on a specific feature of a specific part}'' (OS18). 
Current CAD platforms do not support semantic search across version histories; search functions are limited to rudimentary keyword matching in system-generated attributes (in \textsc{Onshape}~\cite{onshape_help_versions_nodate} and \textsc{SolidWorks}~\cite{dassault_systemes_searching_nodate}) or user-authored documentation (in \textsc{SolidWorks}~\cite{dassault_systemes_searching_nodate}). 

\subsubsection{Editing Version History}
\label{sec:editable_history}

Our analysis revealed that the common cluttered version history, filled with minor or unintended updates, not only creates navigation challenges (Section~\ref{sec:navigate_ver}) but also fuels users' desire to retroactively edit or prune the version history, altering its immutability. This issue was mentioned in 45 forum threads. 
Cloud-based CAD platforms exacerbate this issue by automatically generating new versions to synchronize data and prevent data loss, either periodically in \textsc{Autodesk Fusion}~\cite{autodesk_fusion_milestones_nodate} or continuously to enable DVCS in \textsc{Onshape}~\cite{baran_under_2015}. While this automation ensures reliability, it also floods the version timeline with entities that do not reflect meaningful design changes. One user described this as ``\emph{the equivalent of both extreme hand holding and the software behaving like IT is in charge and not you the creator and designer}'' (AD15), pointing to tensions between data integrity and user autonomy. 
Moreover, users often create versions unintentionally that do not include meaningful design updates, further cluttering the version history. For example, a user may ``\emph{mistakenly save a file with a section analysis visible, changing it to unvisible acts as a major change and once saved, the version number increases}'' (AD18). 

While version control ensures traceability and avoids editing conflicts between collaborators, the accumulation of insignificant versions complicates both review and communication of design evolution, and can also introduce regulatory and compliance risks.
As a user explained: ``\emph{In our regulated industry, we normally have to engage the regulators on any product change and also gain their approval on changes. This is a long process that can take weeks or months, so we cannot afford to have phantom changes propagating up through the product structure that the regulator does not understand since there is no `real' change}'' (OS22). 
This highlights that the impact of noisy version histories extends beyond day-to-day navigation challenges to broader organizational and compliance processes.

Despite these frustrations, CAD platforms enforce immutability of versions once created. Editing versions in CAD carries risks not seen in textual systems like source code: changes made to an earlier point in the timeline can invalidate parametric constraints to downstream geometry, making history editing a structurally fragile operation.
However, even if version deleting cannot be achieved, our analysis of forum discussions reveals users' strong desire for ``\emph{the ability to flag some versions as obsolete with an option to hide them from the version graph}'' (OS16). Yet, no current CAD platforms support such functionality, leaving users with bloated and static records.

\subsubsection{Identifying Changes between Versions}
\label{sec:identify_changes}

\begin{figure*}
    \centering
    \includegraphics[width=1\linewidth]{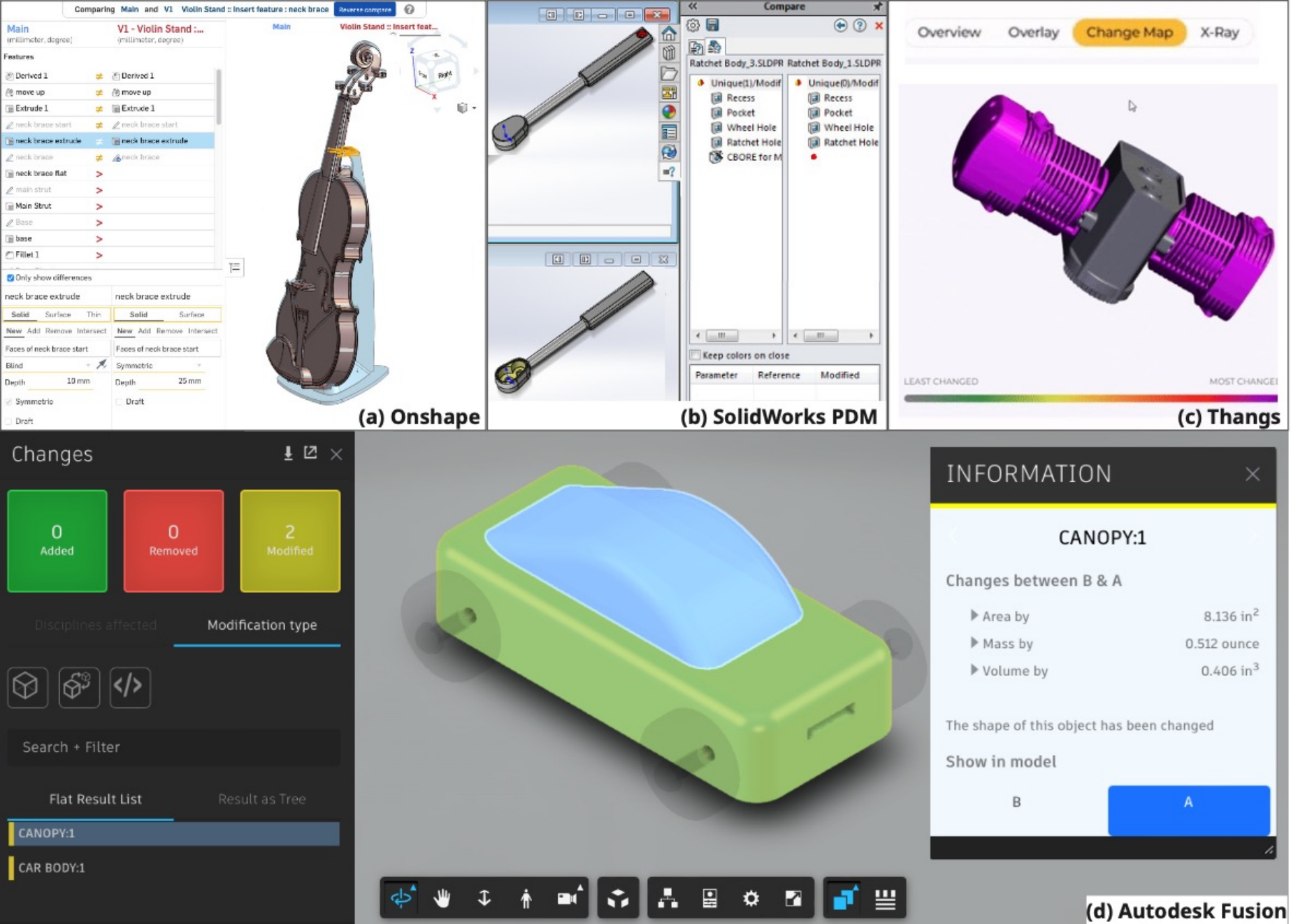}
    \caption{User interface for comparing two versions of a CAD model in the three studied CAD software and \textsc{Thangs}. 
    (a) \textsc{Onshape} presents two model versions with one version overlaid on top of another, where the geometric difference is highlighted in red. All modified modelling commands are listed in the side panel, where selecting each command presents all the modified command parameters, and the geometries that are modified by the selected command are highlighted in yellow. 
    (b) \textsc{SolidWorks} presents two versions of the model side by side, where changes of existing geometries are highlighted in yellow, and additions and deletions of geometries are highlighted in red. Modelling commands that were modified are also listed in the side panel, where selecting each command presents all the modified command parameters (adapted from Ref.~\cite{goengineer_comparing_2017}). 
    (c) \textsc{Thangs} presents an overlaid view of two uploaded model versions, with geometric difference highlighted in purple (adapted from Ref.~\cite{thangs_3d_thangs_2022}). 
    (d) \textsc{Autodesk Fusion} presents only one version of the model that is under comparison with a list of modelling commands that are modified. However, selecting each listed modelling command only highlights the geometries that are modified by the selected command, but not the modified parameters. Modelling command parameters are not compared, even if they are accessible from the uploaded file format. 
    All screenshots of software interfaces are presented for comparison purposes only.}
    \Description{This figure presents the presentation of version comparison results in the three commercial CAD modelling platforms and a CAD sharing platform Thangs as a comparison. Onshape and SolidWorks present version comparison results inside the modelling space with parametric and geometric differences highlighted on the model itself. Thangs and Autodesk Fusion present version comparison results as a new window that visualize geometric differences only. Key characteristics of the visualization are summarized in the figure caption.}
    \label{fig:version_compare}
\end{figure*}

When revisiting a CAD file's version history, users often need to understand the \emph{semantic} differences between versions, beyond geometric changes, which is a concern mentioned in 17 forum threads. As a user pointed out: ``\emph{Due to the nature of CAD data (unlike software code), it is not easy to do a simple `compare' to figure out changes and why. Such visual compare tools (in CAD but not so much in PDM) show `what' changed. But users still have no way to know `why' the changes were made, either in CAD or in PDM}'' (CF23). 
The same user further explained that ``\emph{the `why' and `what' are really the ECOs/ECNs for the part}'' (CF23), where the engineering change orders (ECOs) or engineering change notices (ECNs) are part of the formal process of documenting engineering changes (the ``what'') and the rationales behind making these changes (the ``why'')~\cite{jarratt_engineering_2011, wright_review_1997}. 
As shown in Figure~\ref{fig:version_compare}a, b, and d, all three studied CAD model development platforms support geometry-level version comparison only. These tools list parametric edits of the modelling commands and visualize geometric differences, but offer limited support for higher-level insights such as intent or downstream implications. Interpreting these differences remains labour-intensive and error-prone, especially when changes are subtle or cumulative. 

The challenge is exacerbated when files are exported and shared on platforms like \textsc{Thingiverse} and \textsc{Thangs}. Exported files often use proprietary or neutral exchange formats that discard the version history and the modelling operations that constructed the model (i.e., dumb geometry that loses editability). 
As one user noted, these files are ``\emph{all binary stuff, so it's impossible to diff}'' (RD22). Although \textsc{Thangs} recently introduced visual comparison of geometric differences between versions, as shown in Figure~\ref{fig:version_compare}c~\cite{thangs_3d_thangs_2022}, this feature still requires manual interpretation. 
In practice, users often rely on inconsistent user-generated descriptions to interpret changes, which vary widely in quality~\cite{cheng_lot_2024}.
As a result, users need to constantly reconstruct the reasoning behind changes by manually examining all design parameters, which is an increasingly time-consuming process for large and interdependent models. 

\subsubsection{Branching and Merging Versions}
\label{sec:branch_merge}

As reviewed in Section~\ref{sec:bkg_branch}, prior work has noted persistent shortcomings in branching workflows for CAD systems~\cite{cheng_user_2023}. Our findings echo these concerns and highlight similar limitations in merging design branches, mentioned in 10 forum threads. 
Users frequently compared their expectations to software VCS, hoping for similar flexibility in CAD, such as ``\emph{a GitHub-like version control where you can create and fork different designs off of a base design and also merge back together different design branches}'' (AD15). 

In practice, however, merging remains technically challenging for CAD platforms like \textsc{Autodesk Fusion}.
Unlike text-based code, CAD models encode interdependent geometric and parametric data, making reconciliation across branches difficult. Yet users still expect comparable flexibility and manageability in managing branches offered software VCS like \textsc{Git}. 
As one \textsc{SolidWorks} user put it, ``\emph{the feature-branch workflow with the audit/review step associated to a pull-request is a must. But they're completely useless with CAD files ... (no merging capability)}'' (SW13). 

These experiences highlight a gap between users’ expectations, shaped by modern DVCS tools in software engineering, and the current technical constraints of CAD modelling, where interdependent geometric structures and modelling commands introduce complexities that make automatic merging far more difficult than in software contexts. As a result, while branching is sometimes supported, meaningful merge operations remain limited or absent across mainstream CAD platforms.

\subsection{Continuity of Version Control in CAD}

Version control in CAD needs to handle geometric data that are far more complex than the line-level textual edits of software development. Maintaining continuity across shared, linked, or modified models poses significant technical challenges. 

\subsubsection{Exporting Files with Version History}
\label{sec:history_export}

A recurring concern discovered in the forum threads is the loss of version history once a CAD file is exported or moved outside its original VCS-enabled workspace, mentioned in 19 forum threads. 
Unlike textual code that can be easily cloned and forked while preserving its full version history, modern CAD platforms only export the latest version of a model, where its prior versions become permanently unrecoverable, even if the model is later re-imported. 
This limitation restricts flexibility in file sharing and distribution of work. As one user noted, if ``\emph{someone creates a design in their own project (rather than the parent one for the company), [they will have] issues getting the design moved over}'' (AD16). 
Such system rigidity effectively forces teams to work within a single shared workspace, undermining the distributed collaboration that VCS is intended to support. 

At the same time, re-importing files further compounds the problem, as the version context is often broken: ``\emph{When multiple exported designs reference the same parts, those parts are imported multiple times and references are lost}'' (AD16). 
Since such references are not preserved across export-import cycles, teams much reconstruct these relationships manually. 
This lack of portability not only reduces design fluidity but also undermines collaborative workflows and introduces duplicated work, especially across organizational boundaries and when model components are designed in multiple CAD platforms. 

\subsubsection{Propagating Changes to Linked Files}
\label{sec:change_to_link}

When dependencies are established between a base model and its variants (as described in Section~\ref{sec:bkg_branch}), or when components are integrated into assemblies of parts, the task of propagating updates from the source model to all linked files remains a largely manual and error-prone process. This challenge was mentioned in 26 forum threads. 
As an \textsc{Onshape} user described: ``\emph{Every modification I make (to a part [...] in a separate document) requires a new version before I can see the changes in the assembly. This is very time consuming}'' (OS20). 
After version creation, users also need to manually update every file that the modified part is used in (e.g., an assembly). This workflow can be especially confusing for beginners, because ``\emph{[the users] are not updating the [assembly itself], but the part `to' the already updated [assembly]}'' (OS19). 

While some users appreciated this deliberate control so that they ``\emph{don't have to worry about assemblies changing when [they]'ve made part changes that aren't approved and released}'' (OS19), others called for streamlined mechanisms, such as ``\emph{a simple button to push on the menu bar that will update the parts automatically, without having to select them all individually}'' (OS19). 
These accounts illustrate the trade-off between maintaining deliberate control over design stability and approval workflow, and the need to reduce the manual overhead of keeping linked files synchronized. As designs grow in scale and complexity, and as teams vary in expertise and organizational practices, these competing needs become increasingly difficult to reconcile through current systems. 

\subsubsection{Updating Changes to the Source File}
\label{sec:change_to_source}

While Section~\ref{sec:change_to_link} discussed propagating updates from a source model to its linked files, users also reported challenges in pushing changes made within a linked file back to the source model, mentioned in 11 forum threads. 
When components are inserted into an assembly, users often prefer to modify them \emph{in-context} (i.e., within the assembly where the part model is inserted), instead of editing the source file. 
As one user explained: ``\emph{I want to keep all of my components and assemblies matched. If I update it within the assembly I would like for it to update component parts. Or if I update it within the base components I would like it to update in the assembly}'' (AD15). 
Although in-context editing is now supported in all studied CAD modelling platforms, the workflow still requires users to ``\emph{create a version on every edit to see these updates in the assembly}'' (OS21). Instead, an \textsc{Onshape} user envisioned a more fluid workflow: ``\emph{My desired behavior would be to insert the part as the unversioned top of the branch so changes are reflected in real time, then apply a version once I'm finished}'' (OS21). 
Such requests highlight users' desire for workflows that preserve the benefits of versioning while supporting more fluid, real-time propagation of edits across assemblies and their source components. 

\subsection{Scope of Version Control in CAD}

Version control preserves design history, yet CAD platforms differ in what can be versioned and at what granularity, ranging from flexible tracking to rigid, system-imposed structures. These constraints often reflect necessary trade-offs that are not present in software engineering: given the complexity and interdependence of CAD models, CAD systems impose limitations to maintain stability and performance. 

\subsubsection{Versioning at Greater Granular Level}
\label{sec:version_granularity}

\begin{figure*}
    \centering
    \includegraphics[width=0.7\linewidth]{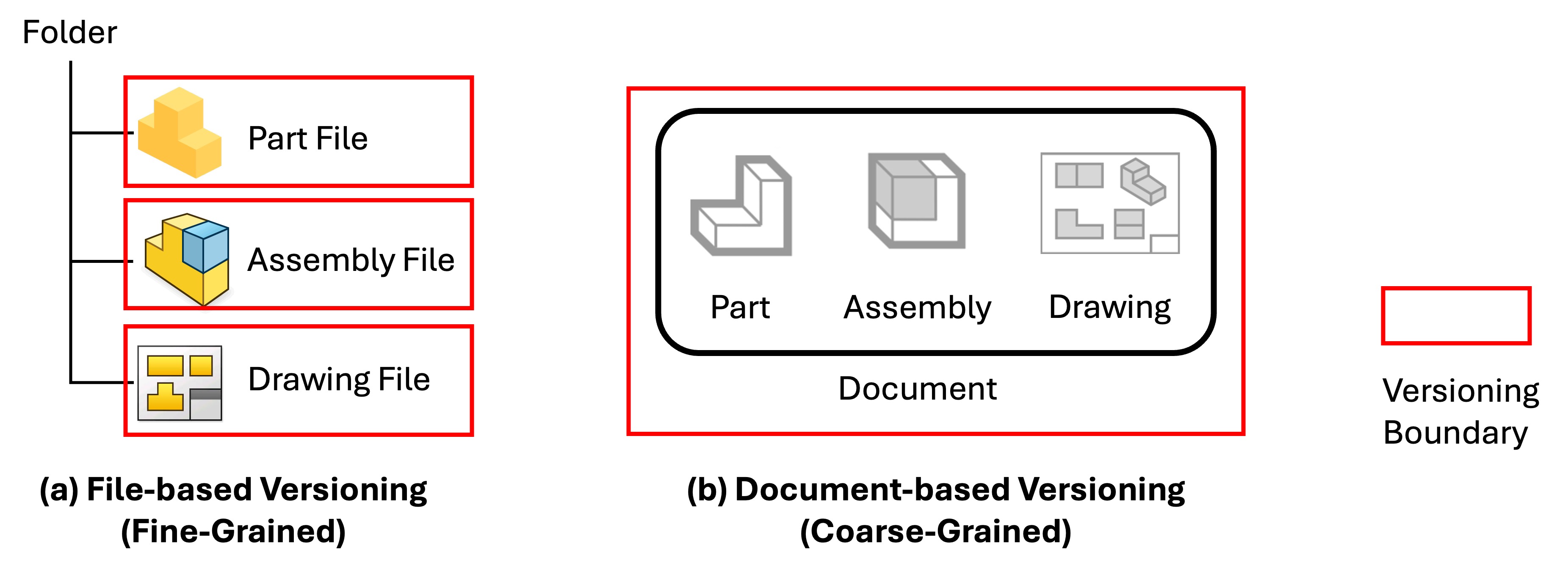}
    \caption{Comparison of versioning granularity across CAD platforms. 
    (a) In file-based platforms like \textsc{SolidWorks} and \textsc{Autodesk Fusion}, each part, assembly, and drawing exists as a separate file, and changes are tracked independently at the file level. 
    (b) In \textsc{Onshape}, all components are grouped into a single document, and versioning applies to the entire document, where changes made in any component could create a new version for the whole document.}
    \Description{This schematic representation describes the difference between file-based CAD platforms and Onshape in versioning granularity. File-based CAD platforms follow a traditional folder structure, where multiple part, assembly, and drawing files may be stored under the same folder directory, and every one of these file is an individual versioning unit with its own versioning boundary. In document-based versioning system such as Onshape, the versioning control is more coarse-grained, where every document contains multiple parts, assemblies, and drawings, and the versioning boundary surrounds the entire document as one global versioning unit.}
    \label{fig:file_structure}
\end{figure*}

While \textsc{Onshape} enables DVCS for CAD modelling, it organizes and versions data at a coarser granularity, which limits users from modular editing. This concern was mentioned in 21 forum threads. 
As shown in Figure~\ref{fig:file_structure}, \textsc{Onshape} restricts versioning to the document level, where a single document may contain multiple parts, assemblies, and drawings. In contrast, traditional file-based systems such as \textsc{SolidWorks} and \textsc{Autodesk Fusion} allow each of these files to be versioned independently. 
Users described this document-level structure as creating unnecessary versioning overhead, compounding the challenge of propagating changes to linked files (Section~\ref{sec:change_to_link}). For example, when users build ``\emph{a `library' document for things like bolts, brackets, etc., and a new item is added to that library. It will cause an unnecessary version bump which can trigger a `newer version' update request in a bunch of other documents where nothing has actually changed}'' (OS16). 
In large CAD documents, this limitation further exacerbates other earlier challenges. For instance, identifying small localized changes becomes more difficult (Section~\ref{sec:identify_changes}), and version histories become increasingly cluttered with minor model updates (Section~\ref{sec:navigate_ver}). 

As a workaround, some users attempted to approximate file-level granularity by splitting large designs across multiple documents. Yet, this approach undermined \textsc{Onshape}'s advantages: ``\emph{With the current setup, reverting to a previous [version of individual parts] would only be useful if I used a new document for every part, but this counteracts the convenience of creating multiple parts}'' (OS18). 
Still, not all users viewed this constraint negatively. As one user reflected: ``\emph{From a data perspective, [this is] a small cost for the advantage of a very powerful tool}'' (OS19).
While some value precise control over individual components, others prefer the simplicity and cohesion of a unified workspace, even at the expense of granularity.

\subsubsection{Untracking Unused Parts}
\label{sec:break_unused_version}

In \textsc{Autodesk Fusion}, users cannot delete any component that was once referenced in a past version of another file (e.g., an assembly that previously contained it), limiting users' ability to clean up and organize workspaces over time. This concern was raised in 12 forum threads. 
This restriction preserves version integrity, ensuring earlier versions can be restored without broken references. However, as projects grow, assemblies evolve and certain parts become obsolete, and ``\emph{the ability to delete past designs and therefore old stale references [becomes crucial]. Meanwhile, [users'] design browsers are just landfills}'' (AD15). 
In the absence of more flexible cleanup mechanisms, some users resort to improvised workarounds, such as ``\emph{placing [the unused files] in a bad files folder}'' (AD16). While effective in the short term, these practices increase project complexity and waste storage space over time.

\subsection{Distribution of Version Control in CAD}

In collaborative CAD environments, cloud platforms distribute version histories across team members, enabling shared access to evolving designs. While this supports coordination, users raised concerns about lacking fine-grained control over what aspects of the version history are shared and with whom.

\subsubsection{Controlling User Access to Versions}
\label{sec:version_access}

When CAD models are shared with external collaborators, document owners often lack fine-grained control over the version history access, a concern mentioned in seven forum threads. 
In \textsc{Onshape}, sharing a project reveals the entire version history~\cite{onshape_help_share_2025}, whereas at the other extreme, \textsc{Autodesk Fusion} limits sharing access to a single version only~\cite{fusion_help_share_2025}. 
This limitation presents practical and strategic challenges. For instance, a user explained that: ``\emph{I am trying to share a version of a document to an outside vendor for quoting, along with download privileges so they can evaluate in their own CAD tool if needed. But I do not want the vendor to have access to the real-time document, only the `quote version'}'' (OS18). 

To achieve more fine-grained sharing control, some users resort to cumbersome workarounds such as: ``\emph{deriving the version of your part studio into a fresh document, or bringing the version of your main assembly into a blank assembly of a fresh document as a sub-assembly [...] Then share that fresh document to the vendor}'' (OS21). 
Even then, this approach remains flawed: ``\emph{A lot of our quote processes are multi-round with tweaks made between rounds, so creating separate documents for each version would get out of hand}'' (OS21). 
Such user experiences highlight the need for scoped sharing, where version visibility can be tailored to the recipient’s role and the purpose of interaction.

\subsection{Summary of Cross-Platform Challenges}
\label{sec:results_summary}

Across the four thematic areas (i.e., management, continuity, scope, and distribution), we observed that certain challenges are universal across all three studied CAD platforms, while others are platform-specific. Table~\ref{tab:evaluation} synthesizes these findings, highlighting which user needs remain unmet and which challenges are partially addressed or fully supported. 
Although the three CAD platforms examined in this study, and many others not included, offer broadly similar technical capabilities for basic CAD modelling, they differ in their collaborative support mechanisms, including the critically important version control infrastructure. 

In practice, users and organizations adopt and migrate between different CAD systems for diverse reasons, such as functional requirements (e.g., integration with downstream manufacturing tools), economical considerations (e.g., free or subsidized offerings for startups), industry standards (e.g., standardized format in exchanging off-the-shelf parts), operational considerations (e.g., operating system compatibility), and so on~\cite{kannan_multi-criteria_2008}. 
As engineers are trained and gain access to different CAD systems in both academic and professional settings, they experience the varying degrees of collaborative support these tools provide and discover a nuanced and uneven landscape of features, where features well-supported in one system may be absent or less refined in another. 
A prominent example is the DVCS architecture in \textsc{Onshape}, which enables real-time collaboration but shifts version management from the file level to the document level. As a result, users transitioning from traditional, file-based CAD platforms (e.g., \textsc{SolidWorks}, \textsc{Autodesk Fusion}) can struggle to adapt to this coarser granularity of versioning in \textsc{Onshape}, where individual parts and assemblies cannot be independently versioned (Section~\ref{sec:version_granularity}).

\begin{table*}[ht]
    \centering
    \caption{User challenges in CAD version control and the extent of support across studied commercial CAD platforms.}
    \begin{tabular}{l l l *{3}{c} l}
        \toprule  
        \textbf{Challenge} & \textbf{Sec.} & \textbf{$n$} & \textbf{\textsc{SolidWorks}} & \textbf{\textsc{Fusion}} & \textbf{\textsc{Onshape}} & \textbf{Challenge type} \\
        \midrule  
        \textbf{\emph{Management of version control}} \\ 
        Navigating version history & \ref{sec:navigate_ver} & 47 & \HalfCircleLeft & \HalfCircleLeft & \HalfCircleLeft & Universal \\
        Editing version history & \ref{sec:editable_history} & 45 & \CircleShadow & \CircleShadow & \CircleShadow & Universal \\
        Identifying changes between versions & \ref{sec:identify_changes} & 17 & \HalfCircleLeft & \HalfCircleLeft & \HalfCircleLeft & Universal \\
        Branching and merging versions & \ref{sec:branch_merge} & 10 & \CircleSolid & \CircleShadow & \CircleSolid & Platform specific \\ 
        \midrule  
        \textbf{\emph{Continuity of version control}} \\ 
        Exporting files with version history & \ref{sec:history_export} & 19 & \CircleShadow & \CircleShadow & \CircleShadow & Universal \\
        Propagating changes to linked files & \ref{sec:change_to_link} & 26 & \CircleSolid & \HalfCircleLeft & \HalfCircleLeft & Platform specific \\
        Updating changes to the source file & \ref{sec:change_to_source} & 11 & \CircleSolid & \CircleSolid & \CircleSolid & Addressed \\
        \midrule  
        \textbf{\emph{Scope of version control}} \\ 
        Versioning at greater granular level & \ref{sec:version_granularity} & 21 & \CircleSolid & \CircleSolid & \CircleShadow & Platform specific \\
        Untracking unused parts & \ref{sec:break_unused_version} & 12 & \CircleSolid & \CircleShadow & \CircleSolid & Platform specific \\
        \midrule  
        \textbf{\emph{Distribution of version control}} \\ 
        Controlling user access to versions & \ref{sec:version_access} & 7 & \CircleSolid & \HalfCircleLeft & \HalfCircleLeft & Platform specific \\
        \bottomrule 
    \end{tabular}
    {\footnotesize
    \begin{tabular}{@{} p{0.8\linewidth} @{} @{} @{} @{}}
        \textbf{Note}: \CircleSolid \space denotes that the challenge is fully resolved with all user needs observed in this study supported, \CircleShadow \space denotes that the challenge is currently unaddressed, and \HalfCircleLeft \space denotes that the challenge is partially resolved but with unmet user needs reported in this study for the specific CAD platform, as of August 2025.
    \end{tabular}} 
    \label{tab:evaluation}
\end{table*}

\section{Design Opportunities}
\label{sec:design_opportunities}

CAD has long served as a cornerstone technology in product development. Over the decades, its role has evolved from a standalone design tool to a complex socio-technical system that mediates collaboration among teams and organizations and supports the creation of increasingly intricate design artifacts.  
To sustain this shift toward large-scale, collaborative engineering design, version control has become a critical component of the underpinning \emph{information infrastructure} in modern CAD systems. Similar to related fields such as software engineering~\cite{dabbish_social_2012, bird_promises_2009} and scientific collaboration~\cite{lu_collaborative_2011, olechnicka_geography_2018, sonnenwald_scientific_2007, hara_emerging_2003, correia_effect_2019}, version control in CAD constitutes the backbone of coordination work. 
In line with Star and Ruhleder's conception of infrastructure~\cite{star_steps_1996}, version control operates largely in the background, quietly supporting collaboration, yet it both shapes and is shaped by user practices. More importantly, such infrastructures often becomes visible only when they break down, revealing the complexities of the collaborative work they support~\cite{star_steps_1996, hanseth_designing_2001}. 
Building on the empirical user challenges identified in our study, this section discusses new research directions and design opportunities aimed at improving the support for collaborative design in CAD modelling platforms with version control. 

\subsection{Supporting the Articulation Work of the Design Process}
\label{sec:opportunity_articulation}

\parabox{
\textbf{Opportunity 1.} During the product design process, users often create new model versions with inconsistent or incomplete change summaries (Sections~\ref{sec:navigate_ver} and~\ref{sec:identify_changes}). As versions accumulate, it becomes increasingly challenging to navigate version histories (Section~\ref{sec:navigate_ver}), with significant design milestones obscured by minor updates (Section~\ref{sec:editable_history}). 
These challenges suggest an opportunity to design tools that better support users in articulating the design history as a collaborative record. Specifically, such tools should (1) help users more consistently capture and describe version differences, (2) support users to organize and edit existing versions to reflect the evolving design process, and (3) present version histories in ways that are communicable to collaborators.
}

Throughout a product's evolution, continuous effort is required to manage the design process as a shared collaborative record, one that extends beyond collaborators' individual personal information spaces and integrates contributions from all team members (i.e., the articulation work)~\cite{schmidt_taking_1992, boyd_understanding_2017, strauss_articulation_1988}. 
In fact, the ease of processing the history of a product may potentially influence the users' impression of the design and ultimately influence work outcomes~\cite{marlow_effects_2015}. 
As evident in software engineering, version control effectively records the information necessarily and supports mechanisms in constructing a collaborative narrative~\cite{bird_promises_2009}. However, due to the complexity of CAD models, a gap in supporting the articulation work of the design process remains unaddressed. 

\paragraph{Capturing and describing version differences}

When versions are created in CAD, users often label them with consistent but opaque numeric identifiers (Section~\ref{sec:navigate_ver}) or with more descriptive but inconsistent, and potentially incomplete, summaries (Section~\ref{sec:identify_changes}), complicating how design histories are articulated and understood. 
As one CAD user noted in Section~\ref{sec:identify_changes}, effective version comparison should capture both the ``what'' (parametric/geometric differences) and the ``why'' (functional difference). Traditionally, such information is documented through formal engineering change notices or orders (ECNs/ECOs)~\cite{jarratt_engineering_2011, wright_review_1997}. 
While the parametric differences may be algorithmically identified, the design knowledge that users leverage and their design intent are often tedious or challenging to capture~\cite{krosnick_think-aloud_2021}. 
As modern version control infrastructures streamline change procedures to support rapid and flexible iterations, they also reduce the reliance on structured documentation practices, leaving much contextual information unrecorded. 

Rather than replacing the articulation needs, technology support should scaffold articulation work~\cite{mark_supporting_2002}. Without undermining the rapid design iterations enabled by modern CAD systems, there are opportunities to augment human productivity through closely supervised applications of artificial intelligence (AI). 
While existing rule-based comparison methods are already effective at detecting geometric differences effectively when models are compared in proprietary formats~\cite{briere_cote_comparing_2012, briere-cote_3d_2013} or even in neutral formats~\cite{lv_difference_2024}, the most laborious challenge remains the synthesis of both major and subtle design changes into meaningful yet concise documentation. 
Drawing on prior successes in automating code summarization using machine learning (ML) techniques in software engineering~\cite{cortes_coy_automatically_2014, leclair_improved_2020, liu_neural_2018, zhang_retrieval-based_2020, linares-vasquez_changescribe_2015}, future work may explore the use of large language models (LLMs)~\cite{pang_understanding_2025} and vision-language models (VLMs)~\cite{zhang_vision-language_2024} to assist in processing both the underlying code representation of CAD models and the visualized geometric differences, such that candidate textual descriptions of the modifications can be generated. 

Alternatively, extending the concept of think-aloud computing~\cite{krosnick_think-aloud_2021} offers a complementary approach, in which users verbalize their thoughts and observations while inspecting 3D geometric differences, or even during the entire modelling process, and the resulting speech is subsequently edited or potentially further summarized with LLM support into textual documentation. 
Prior work suggests that this approach can capture richer contextual information including the design intent, while reducing manual documentation effort~\cite{krosnick_think-aloud_2021}. 
Across these potential approaches, users must remain accountable for the version history entries they create, and the AI should only function as an assistive layer that enhances productivity without displacing human judgment. Standardized documentation templates may be one promising mechanism for structuring this interaction, simultaneously guiding ML models to structure its output and supporting users in reviewing and refining AI-assisted documentation. 

\paragraph{Organizing and editing version histories}

While immutability ensures traceability~\cite{chou_unifying_1986}, current CAD systems provide limited mechanisms for editing or reorganizing records after the fact, degrading the navigability and interpretability of version histories as meaningless versions accumulate over time (Sections~\ref{sec:navigate_ver} and~\ref{sec:editable_history}). 
In contrast, \textsc{Git}-based workflows in software engineering support retrospective cleanup and collaborative narrative construction, allowing contributors to co-construct a version history that serves not merely as a raw log of edits, but as a curated reflection of collective progress~\cite{bird_promises_2009}.

Although CAD geometries and the interdependence among model components are more complex than text-based source code, CAD systems like \textsc{Onshape} have begun modernizing their version control support from CVCS to DVCS, recording incremental design changes when modelling operations are modified, rather than saving full model copies~\cite{baran_under_2015}. This shift opens opportunities for CAD systems with DVCS to implement more sophisticated \textsc{Git}-like operations that allow users to clean up and annotate version histories without compromising traceability. 
However, for CAD systems that remain CVCS-based and open-source environments where heterogeneous model formats are exchanged, DVCS-style functionality may not yet be technically feasible. In these cases, as users suggested in Section~\ref{sec:editable_history}, lightweight curation tools such as customizable markers or tags could be used to foreground significant milestones and improve the interpretability of version histories while preserving the complete underlying record. 

\paragraph{Presenting version histories}

\begin{figure*}
    \centering
    \includegraphics[width=\linewidth]{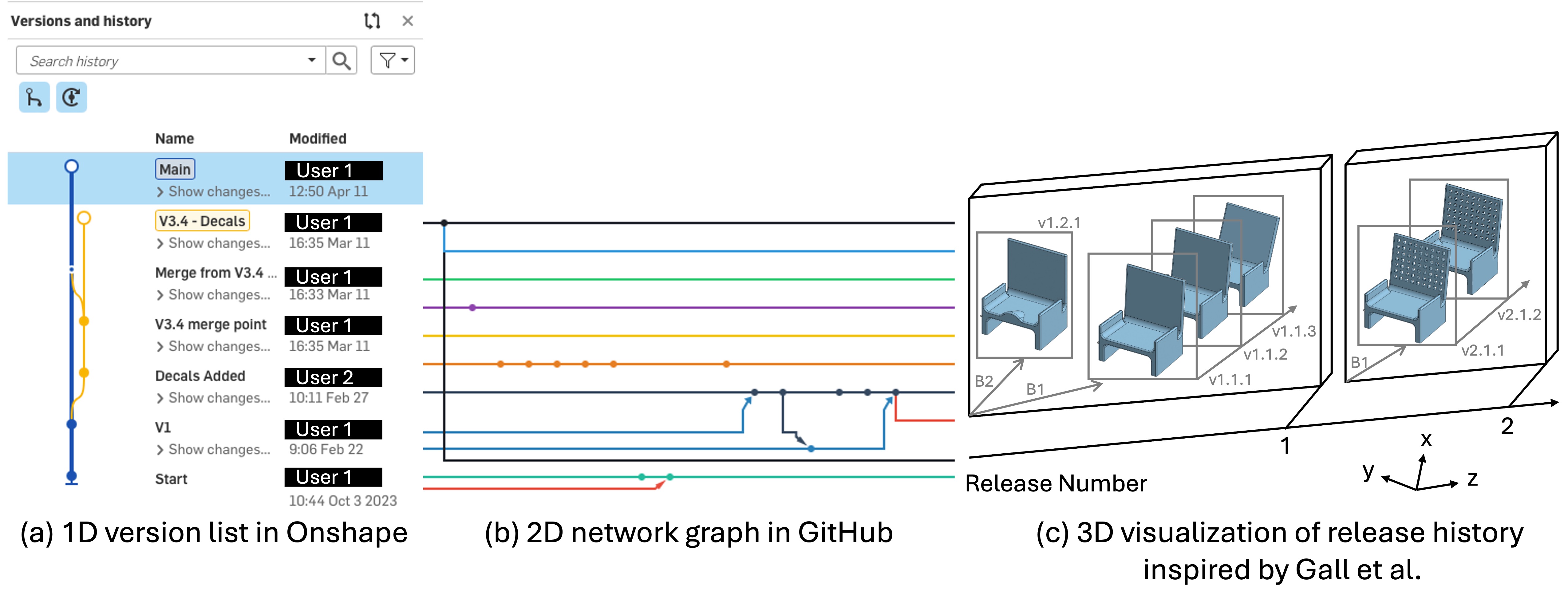}
    \caption{Sample version history visualizations with information presented in various dimensions. 
    (a) \textsc{Onshape}, as a representative of modern commercially available CAD software, presents the version history of a design project with a 1D graph, listing all versions in chronological order. 
    (b) \textsc{GitHub}, a software development platform with DVCS, presents the development history in the network view with branches of versions shown in parallel. 
    (c) Inspired by Gall et al.~\cite{gall_visualizing_1999}, a third dimension can also be utilized to represent the release sequence number (RSN) of a product, where all versioning activities of a specific release are presented in a 2D graph, orthogonal to the direction that RSN increments.
    All screenshots of software interfaces are presented for comparison purposes only.}
    \Description{A comparison between version history presentation with different number of dimensions. Onshape presents the full version history with a 1D scrollable list. GitHub presents the software development process in a 2D graph, where each parallel thread shows a different branch of development. A visualization method proposed by Gall et al. presents the software release history using the 3D space, where every scale unit along the timeline is a 2D graph of the development activities in a specific release. Key characteristics of the visualization are summarized in the figure caption.}
    \label{fig:history_vis}
\end{figure*}

As a design project evolves, its version history inevitably grows, often becoming too extensive for users to comprehend and navigate effectively (Section~\ref{sec:navigate_ver}). 
According to Dourish and Bellotti, effective collaboration support should assist users to construct both (1) a high-level awareness of the \emph{character} of others' actions to help them structure their activities and avoid duplication of work, and (2) a low-level awareness of the \emph{content} of those actions to allow fine-grained shared working and synergistic group behaviours~\cite{dourish_awareness_1992}. 
However, the flat and linear format adopted by most commercial CAD platforms (see Figure~\ref{fig:version_view}a) often fails to reflect these layers of awareness, particularly in collaborative settings where branching, merging, and parallel development are common~\cite{cheng_age_2023}. 
As illustrated in Figure~\ref{fig:history_vis}, CAD systems may draw inspiration from visualization techniques in software engineering~\cite{park_eliph_2017, yoon_visualization_2013, gall_visualizing_1999}, incorporating additional visual dimensions, markers, and colour encoding to represent design evolutions and enhance collaborative awareness within version histories. 

\subsection{Facilitating Cross-Boundary Collaboration}
\label{sec:opportunity_boundary}

\parabox{
\textbf{Opportunity 2.} When users collaborate across multiple boundaries -- whether they be parallel development paths within a project (Section~\ref{sec:branch_merge}), linked projects within a CAD system (Sections~\ref{sec:change_to_link} and~\ref{sec:change_to_source}), or even elements across different CAD systems (Section~\ref{sec:history_export}) -- technical constraints in the modern CAD ecosystem often lead to discontinuities in version control and poor interoperability. At the organizational level, these challenges are compounded by limited support for collaboration across functional roles, including insufficiently granular sharing and access control (Section~\ref{sec:version_access}). 
Together, these challenges highlight opportunities to (1) design mechanisms that support parallel product development and merge reconciliation, (2) design mechanisms that facilitate change propagation across linked projects within a system, (3) develop standards that enable seamless data exchange across proprietary platforms, and (4) design mechanisms that strengthen communication between design teams and external stakeholders. 
}

Boundaries are inherent and fundamental to socio-technical design, supporting effective project management within teams and organizations, yet boundaries should also facilitate the sharing of knowledge and experience~\cite{cherns_principles_1976, mumford_story_2006}. 
Collaborative work around boundary objects often involves negotiating, redefining, and transforming the boundaries~\cite{lee_boundary_2007, carlile_transferring_2004}, and such objects become critical components of the information infrastructure that mediate collaboration across teams and organizations~\cite{gal_dynamics_2008}. 
In our context, each CAD model can operate as a boundary object, where it provides a shared reference through which design intent, manufacturability, and downstream considerations are coordinated~\cite{subrahmanian_boundary_2003}. 
However, as collaboration increasingly expands beyond any single project's predefined boundaries -- across branches, linked projects, organizational roles, and software systems -- our study identifies new forms of friction emerging at multiple boundary levels. 

As summarized in Table~\ref{tab:evaluation} and discussed in Section~\ref{sec:results_summary}, most challenges in cross-boundary collaboration within the same CAD platform appear to be platform-specific. 
Users report tedious and error-prone processes when propagating changes between linked projects (Sections~\ref{sec:change_to_link} and~\ref{sec:change_to_source}), and they experience frustration with limited fine-grained sharing control when collaborating with external partners (Section~\ref{sec:version_access}). 
Similarly, while branching and merging are supported in systems such as \textsc{SolidWorks} and \textsc{Onshape}, they remain limited or absent in \textsc{Autodesk Fusion} and most OSH platforms (Section~\ref{sec:branch_merge}). 
These challenges are not inherent technical constraints of version control infrastructures, but symptoms of uneven infrastructural maturity and diverging product design priorities across the CAD landscape. 
This suggests an opportunity for software providers to better support collaborative workflows by aligning functionality across platforms and recognizing cross-boundary collaboration as a core aspect of design work, rather than a peripheral feature. 
As Schmidt and Simone argued, distributed collaboration relies on constructing and running malleable and linkable computational coordination mechanisms~\cite{schmidt_coordination_1996}, but such mechanisms remain underdeveloped in several CAD systems today. 

In contrast, the challenges of cross-platform collaboration in CAD point to deeper infrastructural fragmentation that demands solutions and standards at the ecosystem level. 
Unlike textual code that can be readily read and edited in any text editor, all CAD platforms operate with different proprietary file formats. When CAD models are exchanged between systems, they are typically exported either in proprietary formats that are not interoperable when imported in other systems, or in standardized neutral formats such as \texttt{STEP}~\cite{iso_step-file_2002} that retain only static geometry while discarding version histories and other metadata~\cite{kasik_ten_2005} (Section~\ref{sec:history_export}). 
Once a model is exported, its associated version history is effectively lost, disrupting the continuity of the boundary object that mediates the shared understanding of design evolution. 
As design traceability and version control become increasingly vital for coordinating distributed work, CAD platforms may seek to build upon existing industry standards to enable bundled exports of multiple selected versions (e.g., milestones, releases) in interoperable formats and accept corresponding bundled imports that restore such data, akin to the ``\verb|git bundle|''\footnote{https://git-scm.com/docs/git-bundle} mechanism in software engineering. 
While these exported models would remain non-editable, such an approach would preserve essential historical checkpoints and capture key design variants, allowing teams to revisit or restore specific design states even when collaborating across heterogeneous CAD environments. 
As the CAD ecosystem matures toward a more standardized and interconnected information infrastructure, it will be better positioned to sustain the continuity of CAD models as boundary objects, and hence support collaborative design work across organizational and technical boundaries. 

From an HCI perspective, these discrepancies in collaboration support at the infrastructure level do not merely motivate calls for ecosystem-level standardization or isolated feature additions, but they also represent a rich site for systems research into how infrastructures are designed, appropriated, and stabilized over time. 
CAD systems are particularly revealing in this regard: once organizations commit to a specific platform, high switching costs (e.g., the manual reconstruction of designs in new proprietary formats) effectively lock in long-term use, allowing the consequences of infrastructural design decisions to unfold across years of collaborative practice. 
As we will further discuss in Section~\ref{sec:opportunity_infrastructure}, users do not simply adapt to infrastructural constraints over time but actively reshape their workflows and develop workaround practices to sustain cross-boundary collaboration. This dynamic creates a distinctive opportunity for HCI systems research to study standards and interoperability mechanisms as socio-technical artifacts, shaped through use and negotiation rather than defined solely by technical specifications. 
As Hanseth and Lundberg argue, complex and domain-specific infrastructures should emerge by bridging communities of practice through standardization~\cite{hanseth_designing_2001}. Rather than directly intervening by proposing a single ``correct'' infrastructure, HCI researchers can leverage established methods to analyze the diverging technical landscape and identify which coordination mechanisms warrant standardization in response to users' evolving collaborative needs. 
In this sense, standards and infrastructure design become an ongoing object of HCI inquiry for supporting cross-boundary collaboration in practice. 

\subsection{Designing Infrastructural Reflexivity}
\label{sec:opportunity_infrastructure}

\parabox{
\textbf{Opportunity 3.} As each CAD system implements version control with distinct technical constraints, users frequently devise their own workarounds to approximate the collaborative workflows that exist in another CAD system (Sections~\ref{sec:version_granularity}, \ref{sec:break_unused_version}, and~\ref{sec:version_access}). 
These practices reveal not merely individual missing features but a broader need for CAD infrastructures that are designed to be reflexive to users and their uses, such that the system is capable of accommodating and learning from the diverse ways they are adapted and appropriated by users in practice~\cite{pipek_infrastructuring_2009}. 
}

Our study reveals how users negotiate infrastructure when the desired system support is lacking. Users creatively develop workarounds, such as organizing obsolete parts in ``bad file folders'' (Section~\ref{sec:break_unused_version}), segmenting documents to approximate finer-grained versioning (Section~\ref{sec:version_granularity}), or duplicating documents to manage version sharing permissions (Section~\ref{sec:version_access}), thereby compensating for platform-level constraints. 
These practices exemplify what Pipek and Wulf describe as \emph{infrastructuring}: the ongoing effort of aligning tools with local needs~\cite{pipek_infrastructuring_2009}. The infrastructural needs are often not recognized until breakdown~\cite{star_steps_1996}. 
We observe that rather than being passive recipients of technical constraints, users actively reconfigure their environments, effectively participating in the construction of their own version control infrastructure. 

Similar to our discussion in Section~\ref{sec:opportunity_boundary}, all the user challenges summarized in this section are also platform dependent. In fact, users often complain about the design workflow in one CAD system simply because they have experienced a more capable workflow in another system. 
This comparative frustration reflects the fact that while major CAD vendors compete to differentiate their products through distinct feature support, the CAD ecosystem lacks a standardized model of VCS. As a result, users must translate their practices across systems on their own to manually bridge the feature gaps. 
A move toward standardized version control as a shared information infrastructure across the CAD ecosystem would require maximizing flexibility, constantly asking whether one solution better supports future specific requirements than another, while recognizing that each solution both enables and constrains actions~\cite{hanseth_developing_1996}. 
As Hanseth and Lundberg note, users will inevitably reshape a new infrastructure during use, and so such infrastructures ``are, and should be, designed and implemented primarily by their users (and in use)''~\cite{hanseth_designing_2001}. 
Taken together, these insights point toward an opportunity to develop CAD infrastructures that evolve with users rather than requiring users to adapt around infrastructural limitations. 

Instead of attempting to anticipate all possible collaborative workflows upfront, we echo the argument that system designers should begin by articulating the principal interactions that the system is meant to support, independent of any particular workflow prescription~\cite{li_beyond_2023, edwards_challenges_2003}. 
Our findings suggest that one such interaction is the ability to treat design changes as something that can be selectively isolated across social and organizational contexts. Designing for this interaction implies defining a minimum yet expressive unit of versioning (e.g., a self-contained part, assembly, or drawing in Figure~\ref{fig:file_structure}) and exposing operations over these units (e.g., versioning, branching, deleting, sharing) as composable primitives rather than fixed workflows. 
While such fine-grained abstractions may introduce technical trade-offs related to performance or security~\cite{edwards_infrastructure_2010}, our study highlights that overly coarse abstractions externalize this complexity onto users, who then reconstruct flexibility through brittle workarounds. 
Designing for infrastructural reflexivity hence shifts the design challenge from predicting use to enabling users to actively shape the infrastructure as part of their ongoing collaborative work. 

\subsection{Reflecting on the Design of VCS in CAD}
\label{sec:design_reflection}

Our study reveals fundamental limitations in how current CAD systems support version control. Rather than being simply another domain to which existing VCS paradigms can be applied, CAD exposes qualities and breakdowns of collaborative infrastructure that remain largely invisible in text-based environments, particularly within an ecosystem shaped by heterogeneous CAD platforms (Section~\ref{sec:opportunity_infrastructure}) and proprietary file formats that lack interoperability (Section~\ref{sec:opportunity_boundary}). 
These challenges reveal forms of articulation work that require collaborators to negotiate design intent through non-textual artifacts (Section~\ref{sec:opportunity_articulation}). 
As cloud-based CAD like \textsc{Onshape} increasingly blurs synchronous and asynchronous modes of work, emerging hybrid collaboration practices challenge the long-standing assumptions about concurrency, traceability, and version lineage in VCS design (Section~\ref{sec:opportunity_articulation}), while simultaneously intensifying inconsistencies across the broader CAD ecosystem (Section~\ref{sec:opportunity_infrastructure}). 
Taken together, these dynamics position version control in CAD as a \emph{work-oriented infrastructure}: an infrastructure that supports highly complex and specialized practices whose properties are largely hidden for those who are not members of the community~\cite{hanseth_designing_2001}. 

While CAD systems can learn from software development in offering more sophisticated VCS mechanisms, the interdependent and semantically rich nature of CAD artifacts require revisiting foundational VCS assumptions. For example, how should incremental changes be defined when enabling DVCS for CAD; how can version differences capture the implicit design intent beyond simple geometric comparison; and how might version histories be designed, managed, and presented to reflect hardware product development workflows. 
These questions signal the need to reconceptualize VCS in CAD as an evolving work-oriented infrastructure shaped through its users and in use, rather than being engineered top-down with assumptions~\cite{hanseth_designing_2001}. 
As CAD vendors develop new versioning features and align functionality across platforms, the socio-technical dynamics in collaboration should remain central, as emphasized in other fields~\cite{edwards_introduction_2009, rawn_understanding_2023, abediniala_facilitating_2022, zhang_vrgit_2023}. 
New features or platform-level constraints should not disrupt the social coordination practices that design teams rely on~\cite{schmidt_taking_1992}. 
Given the absence of a standardized or ``correct'' model for VCS in CAD, future systems should consider flexibility and reflexivity as core design principles, enabling infrastructures that adapt to diverse workflows, supporting ongoing negotiation of meaning, and interoperate both within and across CAD platforms. 

\section{Limitations and Future Work}
\label{sec:limitations}

This paper identifies challenges in adopting version control for CAD-based mechanical product design, with findings shaped by the scope of our data sources. Our analysis is limited by its primary reliance on online forum discussions, which include a significantly greater proportion of issues of model creation in commercial CAD software than challenges of model sharing on open platforms like \textsc{Thingiverse}. 
To further build on these findings, future work may also adopt in-depth qualitative methods (e.g., interviews, contextual inquiry, observational studies) to examine how challenges manifest in different methods of collaboration (e.g., parallel editing, asynchronous collaboration), across industries (e.g., small-scale product design, large-scale engineering design), and within the workarounds users develop to cope with unmet needs. 
Such approaches would provide richer insight into the socio-technical dynamics of collaborative CAD and inform more grounded opportunities for system design. 

Our scope was limited to three commercially available mechanical CAD platforms with built-in PDM solutions, which may not fully capture the diversity of practices across the broader CAD community. 
Other PDM/PLM solutions, such as \textsc{PTC Windchill} and \textsc{Siemens PLM}, support cross-format version control primarily through centralized approaches. While we included \textsc{SolidWorks} in our study to represent this traditional approach, our findings may not generalize to all enterprise workflows. 
At the same time, we acknowledge that the challenges and practices highlighted in this work are not exclusive to mechanical product design but may extend to other domains that also collaborate in comparable CAD environments (e.g., architecture, circuit design). 
Expanding future research to encompass third-party PLM solutions and additional design disciplines would provide a more comprehensive view of professional practices. 

Finally, our analysis foregrounds user challenges while giving less attention to the technical and economic constraints faced by CAD software providers. Unlike source code, CAD data is geometrically complex and interdependent, complicating the direct transfer of \textsc{Git}-like features from software engineering. 
Addressing these challenges will require not only user-centred research but also collaboration with software vendors to understand feasibility and explore viable pathways for innovation.

\section{Conclusion}

In this paper, we presented an empirical study of user-reported challenges in applying version control to modern CAD software. Drawing on data mined from 424 threads across seven online forums, spanning both platform-specific and independent communities, we analyzed 170 relevant discussions to identify common pain points in CAD-based product design related to the management, continuity, scope, and distribution of versions. 
These findings enabled us to reflect on the design of state-of-the-art version control in CAD as an information infrastructure. 
The user challenges surfaced in our analysis further motivated design opportunities for future tools and mechanisms that support articulation work, facilitate cross-boundary collaboration, and operate with infrastructural reflexivity. 
In general, version control in CAD is really the socially-complex, boundary-crossing work that is inevitable in large-scale collaboration for modern product design. 

Our study delivers two sets of implications. 
For CAD software providers, it highlights opportunities to expand current infrastructures to better support collaborative workflows and the distributed, iterative nature of design work. 
For the HCI research community, it positions version control in CAD as a rich site for investigating socio-technical infrastructures, opening avenues for new tools, workflows, and standards to support collaborative product development. 
Ultimately, bridging insights from software engineering and HCI offers a promising path toward re-imagining CAD version control as an information infrastructure that better serves the realities of modern collaborative design practice. 

\begin{acks}
We acknowledge the support of the Government of Canada’s New Frontiers in Research Fund (NFRF) [NFRFE-2022-00543] and the Centre for Analytics and Artificial Intelligence Engineering (CARTE) at the University of Toronto. We are also grateful to our anonymous reviewers for their thoughtful and incisive comments.
\end{acks}

\bibliographystyle{ACM-Reference-Format}
\bibliography{reference}

\end{document}